\documentclass[preprint, amsmath, amssymb, aip, jcp, floatfix]{revtex4-2}
\usepackage{graphicx}
\usepackage{color}
\usepackage{xcolor}
\usepackage{dcolumn}
\usepackage{bm}
\usepackage{amssymb}
\usepackage{hyperref}
\usepackage[utf8]{inputenc}
\usepackage{booktabs} 
\usepackage{siunitx}  
\usepackage{amsmath}   
\usepackage{amsfonts}  
\usepackage{bm}        
\usepackage{mhchem}
\usepackage{float} 
\usepackage{enumitem} 
\newcommand{\ov}[1]{\overline{#1}}

\DeclareMathOperator{\sinc}{sinc}

\begin{document}

\title{Theory of Disorder-Induced Multimode Coupling in a Fabry–Pérot Cavity}

\author{Wei-Kuo Li}
\affiliation{Department of Chemistry and Biochemistry, University of Notre Dame, Indiana 46556, United States}

\author{Hsing-Ta Chen}
\email{hchen25@nd.edu}
\affiliation{Department of Chemistry and Biochemistry, University of Notre Dame, Indiana 46556, United States}

\date{\today}

\begin{abstract}
Strong light--matter interactions have attracted significant interest in physical chemistry due to the potential for modifying molecular processes by hybridizing with cavity photons and forming polaritonic states. 
However, for a disordered molecular ensemble within an optical cavity capable of supporting multiple photon modes, theoretical understanding of multimode polaritons remains incomplete, particularly regarding whether and how distinct photon modes are mutually coupled through molecules. 
In this work, we generalize the collective mode expansion formalism to incorporate multiple photonic modes and show that spatial and orientational disorder in the light--matter coupling strength (originating from dipole orientation fluctuations and photonic field inhomogeneity) can mediate cross-mode coupling. This disorder-induced coupling allows for energy transfer to otherwise unexcited photon modes through a mechanism termed disorder-induced multimode activation (DIMA). 
Using an effective one-dimensional Fabry--P\'{e}rot cavity as a test system, we demonstrate that the cross-mode energy transfer through DIMA can be controlled by changing the molecular layer thickness and the optical properties of the mirrors.
Furthermore, we show that DIMA is robust against weak energetic disorder. This mechanistic understanding lays the foundation for a systematic control of multimode photon coupling using disorder-induced effects.

\end{abstract}

\maketitle
 
\section{Introduction}
Strong light–matter interactions in optical cavities have emerged as a central topic across multiple research areas, including nanophotonics,\cite{gonzalez2024light} quantum plasmonics,\cite{gonzalez2024light,morton_theoretical_2011} and polaritonic chemistry.\cite{ribeiro2018polariton,xiang_intermolecular_2020} The key idea is to leverage coherent exchange between material excitations and photon energy to generate hybrid light–matter quasiparticles, known as polaritons. Polaritons combine photonic and excitonic characters and lead to a characteristic Rabi splitting in optical spectra, which represents a spectroscopic hallmark of strong light–matter interaction. Owing to their hybrid nature, these quasiparticles can redistribute excitation energy and support propagation over extended length scales, suggesting a possible route to harness photonic environments to modify chemistry and material properties.\cite{xu_ultrafast_2023,xiang_molecular_2024,ebbesen_hybrid_2016,xiong_molecular_2023}

In realistic systems, disorder arising from molecular inhomogeneity, dynamical fluctuations, and photonic field distributions is inevitable.
The conventional Tavis--Cummings (TC) model provides a minimal model incorporating $N$ identical two-level systems symmetrically coupled to a single-mode quantized photon field,\cite{tavis_exact_1968} which captures the upper and lower polariton states and $N-1$ optically dark states that does not involve photon excitation. Based on the TC model, disorder can be incorporated as variations in molecular excitation frequencies and in the strength of the light–matter coupling. To account for these disorder effects, perturbative treatments usually introduce random stochastic variables to molecular excitation frequencies\cite{houdre_vacuum-field_1996,zhou_interplay_2022,engelhardt2023polariton} or include Holstein-type vibrational degrees of freedom,\cite{ying_microscopic_2023,ying_resonance_2024,ying_microscopic_2025} leading to polariton localization\cite{engelhardt2023polariton} and mixing of dark and polaritonic states.\cite{catuto2025interplay} 
These perturbation-based methods are justified for treating frequency disorder because frequency disorder is usually small compared to the collective coupling.
However, the light-matter coupling strength of individual molecules can vary significantly due to molecular orientation and the local photonic fields,\cite{mandal2023microscopic,fregoni_theoretical_2022} causing strong disorder that cannot be treated as perturbations. To capture this strong disorder, we develop the collective mode expansion method and show that coupling disorder can induce splitting in linear spectra, which resembles the Rabi splitting as induced by collective coupling.\cite{li_disorder-induced_2025} 

Another missing part of the TC model is that it assumes the cavity includes only a single photon mode, while a realistic cavity often supports multiple photon modes.
Many platforms for strong light-matter coupling, such as confocal cavities and plasmonic metasurfaces,\cite{wickenbrock_collective_2013,wright_versatile_2023,ribeiro_multimode_2022,crookes_collective_2025,reshef_multiresonant_2019,huang_microcavity-assisted_2024,kristensen_modes_2014} can support multiple photon modes that are degenerate or have small mode spacing comparable to Rabi splitting, leading to a collection of multimode polaritons. 
That being said, it remains unclear whether these multimode polaritons are coupled through the molecules or completely decoupled. 
On the one hand, inter-mode energy transfer as induced by intrinsic molecular nonlinearity can be observed by cross-peaks in two-dimensional infrared (2D-IR) spectra of polaritons.\cite{yang2023enabling}
On the other hand, angle-resolved reflectivity data show that, as light-matter coupling strength increases, the reflectivity spectra are fitted to a decoupled polariton $(2M)\times (2M)$ model (where $M$ is the number of the photon modes), rather than a coupled polariton $(M+1)\times (M+1)$ model, suggesting that these polaritons do not exhibit cross-mode energy transfer.\cite{georgiou2021observation}. 
This decoupling transition results from the interplay between the dissipation and delocalization of molecular polaritons within the cavity,\cite{balasubrahmaniyam_coupling_2021} which plays an important role in quantum information processing applications.\cite{prado_atom-mediated_2011, wickenbrock_collective_2013}
 
To simulate the dynamics of multimode polaritons within a realistic cavity setting, many theoretical methods have been developed to integrate multiple photon modes into the disordered TC model. 
For multi-layered materials within an effective one-dimensional Fabry-P\'erot cavity, the full dynamics following a microscopic light-matter Hamiltonian can be reduced to an effective matrix equation through a unitary QR decomposition, resulting in multimode polariton dispersions that agree well with experimental spectra.\cite{mandal2023microscopic} This approach is applicable for arbitrary thickness and position within the cavity, and can be generalized to capture exciton-polariton dynamics.\cite{rahmanian_koshkaki_exciton-polariton_2026} However, recent findings indicate that the degree of delocalization of the dark states increases with the number of molecules per layer, implying that a higher-dimensional representation of the cavity photon modes is required.\cite{sun2025exploring}
In the presence of molecular disorder, multimode polariton localization and dispersion can be estimated using the disorder-averaged Green’s function formalism.\cite{engelhardt2023polariton} This method allows for resolving spatially dependent quantities (such as the coherence length) and their interplay with frequency disorder among molecules.
Despite these advances, there remains a gap in understanding how frequency and orientation disorders influence the coupling among multiple cavity photon modes.

To investigate the interplay between disorder effects and multimode polariton, we generalize the collective mode expansion method to incorporate multiple photon modes within a disordered TC model. Specifically, the electric-field distributions associated with these photon modes exhibit distinct spatial dependencies, and we consider both perturbative frequency disorder and strong variation in light-matter coupling due to orientation disorder. We then analyze their cross-mode coupling in terms of the transient dynamics, steady-state spectrum, and integrated intensity.
This article is organized as follows. Section~\ref{sec:Theory} introduces a theoretical framework for disordered molecular ensembles coupled to a multi-mode optical cavity and derives the corresponding equations of motion using the collective mode expansion. Section~\ref{sec:1D_cavity} specifies the system parameters for a one-dimensional Fabry--Pérot cavity with distributed Bragg reflector (DBR) mirrors and analyzes cross-mode coupling arising from correlations among the dark modes of molecular emitters. In Section~\ref{sec:trade-off}, we demonstrate that tuning the cross-mode correlation by modifications of the molecular layer enables potential control over energy transfer between photonic modes. Finally, Section~\ref{sec:conclusion} summarizes our findings and discusses candidate experimental platforms for realizing these effects.



\section{Model and Theory}\label{sec:Theory}

\subsection{Disordered Molecules in a Multi-Mode Cavity}
We consider a multi-mode Tavis--Cummings model consisting of an ensemble of $N$ molecular emitters coupled to a manifold of $M$ cavity photon modes. The total Hamiltonian takes the form $\hat{H}=\hat{H}_\mathrm{M}+\hat{V}_\mathrm{MP}+\hat{H}_\mathrm{P}+\hat{V}_\mathrm{in}$ where $\hat{H}_\mathrm{M}$ and $\hat{H}_\mathrm{P}$ denote the matter Hamiltonian and the photon Hamiltonian, respectively, and $\hat{V}_\mathrm{MP}$ describes their interactions.
The photon Hamiltonian is given by $\hat{H}_\mathrm{P}=\sum_{k=1}^M\hbar\omega_k (\hat{a}_k^\dagger \hat{a}_k+\frac{1}{2})$ where $\hbar\omega_k$ and $\hat{a}_k$ ($\hat{a}_k^\dagger$) denote the photon energy and annihilation (creation) operator of the $k$-th photon mode. For generality, we assume the cavity photon is not perfectly confined and has a decay rate $\Gamma_k$ for the $k$-th mode, thus we define the complex-valued energy as $\hbar\Omega_k=\hbar\omega_k-i\Gamma_k/2$. 
The cavity photon modes can be driven by the external field through
$\hat{V}_\mathrm{in}={\cal E}_\mathrm{in}(t) \sum_kg_k\hat{a}_k^\dagger+c.c.$ where $c.c.$ denotes the complex conjugate.
Here ${\cal E}(t)$ denotes the time-dependent part of the external field and $g_k$ is the coupling strength corresponding to the $k$-th photon mode.

In the context of cavity quantum electrodynamics, the electric field operator at the position $\mathbf{r}$ can be expressed as follows, 
\begin{equation}\label{eq:E-field}
\hat{\mathbf{E}}(\mathbf{r}) =\sum_k \hat{\boldsymbol{\mathcal{E}}}_k(\mathbf{r}) + \hat{\boldsymbol{\mathcal{E}}}_k^\dagger(\mathbf{r})
\end{equation}
Here, the complex electric field operator of the $k$-th mode is written in the form of
$\hat{\boldsymbol{\mathcal{E}}}_k(\mathbf{r}) = \boldsymbol{\mathcal{G}}_k(\mathbf{r}) \hat{a}_k$ where $\boldsymbol{\mathcal{G}}_k(\mathbf{r})$ represents the spatially dependent factor. Note that, for a vacuum cavity within the Coulomb gauge, the spatial dependent factor is given by $\boldsymbol{\mathcal{G}}^\circ_k(\mathbf{r})=i\mathbf{s}_k\sqrt{\frac{\hbar\omega_k}{2\epsilon_0\mathcal{V}}}e^{i\mathbf{k}\cdot\mathbf{r}}$ where $\mathbf{s}_k$ is a unit vector of transverse polarization associated with the wave vector $\mathbf{k}$ ($k=|\mathbf{k}|$), $\epsilon_0$ is the vacuum permittivity, and $\mathcal{V}$ is the effective volume of the cavity.\cite{cohen-tannoudji_photons_1997,li_disorder-induced_2025}

We model the matter subsystem within the cavity as an ensemble of $N$ identical quantum harmonic oscillators described by the matter Hamiltonian $\hat{H}_\mathrm{M}= \sum_{j=1}^N \hbar\Omega_v \hat{c}_j^\dagger \hat{c}_j$. Here, we denote the complex-valued frequency $\hbar\Omega_v=\hbar\omega_v-i\Gamma_v/2$ where $\omega_v$ and $\Gamma_v$ are the transition frequency and relaxation rate of the emitter, respectively. Note that, while the quantum oscillators are assumed to have identical transition frequencies and relaxation rates, each emitter undergoes a distinct light--matter interaction depending on its position ($\mathbf{r}_j$) and transition dipole orientation ($\boldsymbol{\mu}_j$). 

For light-matter coupling, we adopt the electric dipole approximation with random dipolar orientation. The coupling between the confined electric field and the $j$-th molecule located at $\mathbf{r}_j$ with the polarization $\hat{\mathbf{P}}_j$ takes the form of $\hat{V}_\mathrm{MP}=\sum_{j=1}^N\hat{\mathbf{P}}_j \cdot \hat{\mathbf{E}}(\mathbf{r}_j)$.
The polarization of quantum harmonic oscillators is given by $\hat{\mathbf{P}}_j = \boldsymbol{\mu}_j(\hat{c}_j^\dagger+\hat{c}_j)$ where $\boldsymbol{\mu}_j$ is the transition dipole moment.
Assuming the rotating wave approximation, we can use the electric field operator in Eq.~\eqref{eq:E-field} and write the light--matter coupling in the form of
\begin{equation}
\hat{V}
_\mathrm{MP}=\sum_{j=1}^N\sum_{k=1}^Mu_{jk}  \hat{c}_j^\dagger\hat{a}_k  + c.c.
\end{equation}
For simplicity, we define the coupling strength between cavity photon $k$ and molecule $j$ as
\begin{equation} 
u_{jk} \equiv  \boldsymbol{\mu}_j  \cdot  \boldsymbol{\mathcal{G}}_k(\mathbf{r}_j)
\label{Eq:coupling_constant}
\end{equation}
which represents random variables in the light--matter coupling system. 
In a realistic system, the fluctuation of the coupling strength $u_{jk}$ originates from both the transition dipole orientations and the spatial distribution of the electromagnetic fields for mode $k$.

Given the Hamiltonian, the operators for the cavity photons and quantum emitters follow the Heisenberg equation of motion $i\hbar\frac{d}{dt}\hat{a}_k(t)=[\hat{a}_k(t),\hat{H}]$ for $k=1,\cdots,M$ and $i\hbar\frac{d}{dt}\hat{c}_j(t)=[\hat{c}_j(t),\hat{H}]$ for $j=1,\cdots,N$. 
These correspond to $M+N$ linear differential equations, where $N\gg M$ is usually a large number in a realistic system.

\subsection{The Collective Mode Expansion}

To explicitly analyze the effect of the coupling disorder, we employ the collective mode expansion, which categorizes the matter operators into a symmetric mode and multiple dark modes corresponding to distinct fluctuations.\cite{li_disorder-induced_2025}
First, without loss of generality, we express the coupling strength in terms of $u_{jk} = \ov{u}_{k} + \delta u_{jk}$. Here $\ov{u}_{k}=\frac{1}{N}\sum_{j=1}^N u_{jk}$ represents the average coupling of the $k$-th photon mode and $\delta u_{jk}=u_{jk}-\ov{u}_{k}$ denotes the coupling fluctuation, which follows $\sum_{j=1}^N \delta u_{jk}=0$. For each mode, the symmetric coupling is characterized by its average $\ov{u}_{k}$ and  the coupling fluctuation is characterized by its variance $\sigma_{k}^{2}=\frac{1}{N}\sum_{j=1}^{N}|\delta{u}_{jk}|^{2}$. 

Next, we define the collective operators for the matter subsystem: 
\begin{align}
    \hat{B} &=\frac{1}{\sqrt{N}}\sum_{j=1}^N \hat{c}_j \\ 
    \hat{D}_{k} &=\frac{1}{\sigma_{k}\sqrt{N}}\sum_{j=1}^{N}\delta{u}_{jk}^*\hat{c}_{j}
\end{align} 
Here $\hat{B}$ denotes the symmetric mode (which corresponds to the bright state of the matter subsystem under the average light-matter coupling), and $\hat{D}_{k}$ denotes the dark mode with respect to the $k$-th photon mode (which corresponds to the dark state due to the coupling fluctuation).
By design, we include the $1/\sqrt{N}$ factor for the symmetric mode and the $1/\sqrt{N}\sigma_k$ factor for the dark mode to ensure normalization and preserve the dimension of the collective operators. 

Therefore, the time evolution of these collective mode operators follows the equations of motion (EOMs)
\begin{align}
    &\hbar(\Omega_k - i\partial_t)\hat{a}_{k} + \sqrt{N}\bar{u}_k^*\hat{B}+\sqrt{N}\sigma_{k}\hat{D}_{k}=-g_{k}\mathcal{E}(t)\label{eq:photon_EOM_time}\\
    &\hbar(\Omega_v-i\partial_t)\hat{D}_{k}+\sum_{k'=1}^M\xi_{k,k'}\sqrt{N}\sigma_{k'}\hat{a}_{k'}=0\label{eq:dark_EOM_time}\\
    &\hbar(\Omega_v-i\partial_t)\hat{B}+ \sum_{k'=1}\sqrt{N}\bar{u}_{k'}\hat{a}_{k'}=0\label{eq:bright_EOM_time}
\end{align}
for $k=1,\cdots,M$. The correlation between the $k$ and $k'$ modes is quantified using the dimensionless covariance between $\delta u_{jk}$ and $\delta u_{jk'}$:
\begin{equation}\label{Eq:XI_def}
    \xi_{k,k'} = \frac{1}{N\sigma_{k}\sigma_{k'}}\sum_{j=1}^N \delta u_{jk}^* \delta u_{jk'}
\end{equation}
For two uncorrelated modes, $\xi_{kk'}=0$, and for the same mode, $\xi_{kk}=1$. This covariance arises from the non-commuting property between the dark modes (i.e. $[\hat{D}_{k},\hat{D}_{k'}^\dagger]\neq0$ when $k\neq k'$).
In Eq.~\eqref{eq:dark_EOM_time} and Eq.~\eqref{eq:bright_EOM_time}, the symmetric mode and the dark modes are coupled to the cavity photon through the average coupling $\bar{u}_k$ and the variance $\sigma_k$, respectively.
More importantly, the $k$-th dark mode can couple to other cavity photon modes through the inter-mode correlation.
We emphasize that the equations of motion in the collective mode representation (i.e., in terms of $\hat{a}_{k}$, $\hat{D}_{k}$, and $\hat{B}$) follow the effective asymmetric Hamiltonian. This is because the dark modes $\{\hat{D}_k\}$ do not form an orthogonal basis when $\xi_{k,k'}\neq0$. 

As a final remark, the collective mode expansion decreases the number of equations from $M+N$ to $2M+1$ without introducing any approximation, at the cost of yielding an asymmetric Hamiltonian and describing only the collective mode of the matter subsystem, instead of retaining information about each individual subsystem.
Note that, for any given realization of the disordered system, $\hat{a}_{k}$ is fixed, whereas $\hat{D}_k$ may vary, so it is necessary to average over many realizations.

\subsection{Transient Dynamics}\label{subsec:dynamics}
Based on these reduced EOMs in the collective mode representation, we can propagate the dynamics of the expectation values $\langle \hat{a}_k(t)\rangle\equiv a_k(t)$, $\langle \hat{B}(t)\rangle\equiv B(t)$ and $\langle \hat{D}_k(t)\rangle\equiv D_k(t)$ under an arbitrary external field. To systematically categorize these transient dynamics, we consider the following three cases: 
\begin{enumerate}[label=({\Roman*})]
    \item  The symmetric coupling case ($\bar{u}_k\neq0$, $\sigma_k=0$ for all $k$) represents the idealized situation where the cavity photon modes couple to only the symmetric, bright state ($B$) and the coupling fluctuation is ignored. In this case, Eq.~\eqref{eq:dark_EOM_time} becomes decoupled from the other equations, 
    \begin{align}
    &\hbar(\Omega_k - i\partial_t)a_{k}(t) + \sqrt{N}\bar{u}_k^*{B}(t)=-g_{k}\mathcal{E}(t)\\
    &\hbar(\Omega_v-i\partial_t){B}(t)+ \sum_{k'}\sqrt{N}\bar{u}_{k'}{a}_{k'}(t)=0
    \end{align}
    These EOMs suggest that all the photon modes are coupled to the symmetric state of the molecular excitation, which implies the long-wavelength limit.\cite{rokaj_lightmatter_2018} In this limit, the model is reduced to a coupled oscillator model,\cite{georgiou2021observation,mandal2023microscopic} leading to the standard Rabi oscillation between ${a}_k(t)$ and ${B}(t)$.
    \item The uncorrelated fluctuation case ($\bar{u}_k=0$, $\sigma_k\neq0$ with $\xi_{k,k'}=\delta_{kk'}$) represents a more realistic situation where the coupling fluctuations dominate light-matter interactions. In this case, Eq.~\eqref{eq:bright_EOM_time} becomes decoupled from the cavity photons and the dark modes. 
    \begin{align}
    &\hbar(\Omega_k - i\partial_t){a}_{k}(t) +\sqrt{N}\sigma_{k}{D}_{k}(t)=-g_{k}\mathcal{E}(t)\\
    &\hbar(\Omega_v-i\partial_t){D}_{k}(t)+\sqrt{N}\sigma_{k}{a}_{k}(t)=0
    \end{align}
    Since the dark modes are assumed to be uncorrelated ($\xi_{k,k'}=\delta_{kk'}$), the dark modes $\{{D}_k\}$ form an orthogonal basis and the effective Hamiltonian becomes symmetric. Interestingly, the EOMs turn out to take the same form as Case~(I), suggesting that ${a}_k(t)$ and ${D}_k(t)$ follow a similar Rabi-like oscillation.
    \item The correlated fluctuation case ($\bar{u}_k=0$, $\sigma_k\neq0$ with $\xi_{k,k'}\neq\delta_{kk'}$) describes a generalized setting in which the coupling fluctuations exhibit a certain degree of correlation between different modes. The non-vanishing correlation suggests that the coupling between cavity photon modes is mediated by the dark modes. Furthermore, we can derive the integro-differential equation for ${a}_k(t)$ by plugging the formal solution of ${D}_k(t)$ into Eq.~\eqref{eq:photon_EOM_time}:
    \begin{equation}\label{eq:integro-differential}
        \partial_t{a}_k(t)=-i\Omega_k{a}_k-\frac{ig_k}{\hbar}\mathcal{E}_0(t)\nonumber+\frac{1}{\hbar^2}\int_0^tdt'\mathcal{K}(t-t'){a}_{k'}(t')
    \end{equation} 
    where the memory kernel is defined as $\mathcal{K}(t-t')=\sum_{k'}N\xi_{k,k'}\sigma_k\sigma_{k'}e^{-i\Omega_v(t-t')}$. 
    Importantly, this memory kernel implies that the cavity photon modes are effectively coupled to the dark mode through a dynamical process characterized by the second moment of the coupling fluctuations. We emphasize that non-zero correlation terms ($\xi_{k,k'}$) lead to an effective inter-mode coupling dynamically mediated by the dark modes. We refer to this disorder-activated inter-mode coupling as disorder-induced multimode activation (DIMA).
\end{enumerate}

\subsection{Steady-State Amplitude }
With the EOM in the collective mode representation, we can further derive the steady-state amplitude and optical spectrum.
To do so, we consider a continuous-wave external field ${\cal E}(t)={\cal E}_0e^{-i\omega t}$ and employ the steady-state ansatz for the expectation values, $a_k(t)=\ov{a}_ke^{-i\omega t}$, $B(t)=\ov{B}e^{-i\omega t}$ and $D_k(t)=\ov{D}_ke^{-i\omega t}$. The steady-state amplitudes follow a set of linear equations:
\begin{align}
    &\hbar(\Omega_k - \omega)\ov{a}_{k} + \sqrt{N}\ov{u}_k^*\ov{B}+\sqrt{N}\sigma_{k}\ov{D}_{k}=-g_{k}\mathcal{E}_0 \label{eq:Ak_EOM_steady}\\
    &\hbar(\Omega_v-\omega)\ov{D}_{k}+\sum_{k'}\xi_{k,k'}\sqrt{N}\sigma_{k'}\ov{a}_{k'}=0 \label{eq:Dk_EOM_steady}\\
    &\hbar(\Omega_v-\omega)\ov{B} + \sum_{k'}\sqrt{N}\ov{u}_k\ov{a}_k=0\label{eq:bright_EOM_steady}
\end{align}
for $k=1,\cdots,M$. 
The extinction spectrum can be computed in terms of the energy loss rate of the incident field $P(\omega)=-2\omega\text{Im}\sum_k{\cal E}_0^*g_k^* \ov{a}_k$.
For Case~(\emph{I}), if the external field drives only the $k_0$ mode (i.e $g_k=0$ if $k\neq k_0$), taking $\Omega_{k_0}=\Omega_v$ leads the absorption spectrum to exhibit the standard Rabi splitting characteristic of the strong-coupling regime, in which the two peaks are separated by $2\sqrt{N}|\bar{u}_{k_0}|/\hbar$.

We now focus solely on the effects arising from the coupling fluctuations (i.e., $\ov{u}_k=0$, Cases (II) and (III)) and write Eqs.~\eqref{eq:Ak_EOM_steady} and \eqref{eq:Dk_EOM_steady} in the block matrix form
\begin{equation}
\begin{bmatrix}
    \mathbf{W}_a &\boldsymbol{\Sigma}\\
    \boldsymbol{\Xi} & \mathbf{W}_v 
\end{bmatrix}
\begin{bmatrix}
    \bar{\mathbf{a}}\\\bar{\mathbf{D}}
\end{bmatrix}=
\begin{bmatrix}
    {-\mathbf{g}}\\\mathbf{0}
\end{bmatrix}
\end{equation}
Here, $\bar{\mathbf{a}} = [\cdots \bar{a}_k \cdots]^T$, $\bar{\mathbf{D}} = [\cdots \bar{D}_k \cdots]^T$, and ${\mathbf{g}} = [\cdots {g}_k\mathcal{E}_0\cdots]^T$ are vectors of size $M$. $\mathbf{W}_a$, $\mathbf{W}_v$, and $\boldsymbol{\Sigma}$ are $M\times M$ diagonal matrices with the matrix element given by $[\mathbf{W}_a]_{kk'}=\hbar(\Omega_k-\omega)\delta_{kk'}$, $[\mathbf{W}_v]_{kk'}=\hbar(\Omega_v-\omega)\delta_{kk'}$ and $[\boldsymbol{\Sigma}]_{kk'}=\sqrt{N}\sigma_{k'}\delta_{kk'}$. Note that $[\boldsymbol{\Xi}]_{kk'}=\xi_{k,k'}\sqrt{N}\sigma_{k'}$ is associated with the multi-mode correlations and, in general, is not a diagonal matrix. Formally, the steady-state amplitude of the photon ($\bar{\mathbf{a}}$) can be solved by $\bar{\mathbf{a}} = -(\mathbf{W}_a - \boldsymbol{\Sigma} \mathbf{W}_v^{-1} \boldsymbol{\Xi})^{-1} \mathbf{g}$.

To further analyze $\bar{\mathbf{a}}$, we consider Case~(II) as the unperturbed case and treat the off-diagonal part in a perturbation manner.
In the absence of the inter-mode correlations ($\xi_{k,k'}=\delta_{kk'}$), $\boldsymbol{\Xi}$ reduces to the diagonal matrix $\boldsymbol{\Sigma}$.
In this case, the linear equations can be separated into $M$ distinct sets of independent equations, each associated with a specific combination of $\ov{a}_k$ and $\ov{D}_k$. As a result, the expression for $\ov{a}_k$ can be solved exactly
\begin{equation}\label{eq:a_k_case2}
    \ov{a}_k=\frac{g_k \mathcal{E}_0}{\hbar}\left(\frac{\cos^2\theta_k}{\omega-\Omega_{+,k}} + \frac{\sin^2\theta_k}{\omega-\Omega_{-,k}}\right)
\end{equation}
where the eigenmode frequencies are 
\begin{equation}\label{eq:Omega_case2}
    \Omega_{\pm,k}=\frac{\Omega_k + \Omega_v}{2} \pm \sqrt{\left(\frac{\Omega_k - \Omega_v}{2}\right)^2 + \frac{N \sigma_k^2}{\hbar^2}}
\end{equation} and the mixing angle is given by $\tan\theta_k=\frac{2}{\hbar}\frac{\sqrt{N} \sigma_k}{\Omega_k - \Omega_v}$.

For example, we consider the external field drives only the $k_1$ mode (i.e $g_{k_1}=1$, otherwise $g_k=0$ if $k\neq k_1$) and choose the emitter frequency to be resonant with the $k_1$ mode ($\Omega_v=\Omega_{k_1}$). In this example, we have $\cos^2\theta_{k_1}=\sin^2\theta_{k_1}=\frac{1}{2}$ and the expression for $\ov{a}_{k_1}$ follows Eq.~\eqref{eq:a_k_case2}.  Eq.~\eqref{eq:Omega_case2} suggests that the spectral peak splitting is now determined by $\Omega_{+,k_1}-\Omega_{-,k_1}=2\sqrt{N}\sigma_{k_1}/\hbar$. We refer to this single-mode splitting as disorder-induced spectral splitting (DISS).\cite{li_disorder-induced_2025}

\subsection{Perturbation Treatment}
We now consider Case~(III) in Sec.~\ref{subsec:dynamics} and treat the cross-mode correlation ($\xi_{kk'}\neq\delta_{kk'}$) as a perturbation. We now let $\boldsymbol{\Xi}=\boldsymbol{\Sigma}+\lambda\mathbf{X}$, in which $\mathbf{X}$ includes all the off-diagonal couplings of $\boldsymbol{\Xi}$. The steady-state amplitudes are written as $\bar{\mathbf{a}}=\bar{\mathbf{a}}^{(0)}+\lambda\bar{\mathbf{a}}^{(1)}$, and $\bar{\mathbf{D}}=\bar{\mathbf{D}}^{(0)}+\lambda\bar{\mathbf{D}}^{(1)}$. The zeroth-order solution is
\begin{equation}
    \bar{\mathbf{a}}^{(0)} 
    = -\frac{\mathbf{W}_v}{\mathbf{W}_a\mathbf{W}_v - \boldsymbol{\Sigma}^2}\cdot \mathbf{g}
\end{equation} 
Note that the eigenvalues of $\mathbf{W}_a\mathbf{W}_v - \boldsymbol{\Sigma}^2$ are $\Omega_{\pm,k}$ for all $k$. The explicit solution for $\bar{\mathbf{a}}^{(0)}$ recovers $\ov{a}_k$ for Case~(II) in Eq.~\eqref{eq:a_k_case2}.
The first-order correction is given by 
\begin{equation}
    \bar{\mathbf{a}}^{(1)} = -\frac{\boldsymbol{\Sigma}}{\mathbf{W}_a\mathbf{W}_v-\boldsymbol{\Sigma}^2}\cdot \mathbf{X}\cdot \frac{\mathbf{W}_v}{\mathbf{W}_a\mathbf{W}_v - \boldsymbol{\Sigma}^2}\cdot \mathbf{g}
\end{equation}
Formally, both $\bar{\mathbf{a}}^{(0)}$ and $\bar{\mathbf{a}}^{(1)}$ are a linear combination of $\frac{1}{\omega-\Omega_{\pm,k}}$ for all $k$.

For example, we consider the external field drives only the $k_1$ mode (i.e $g_k=0$ if $k\neq k_1$). The unperturbed solution $\bar{\mathbf{a}}^{(0)}$ has only the $\frac{1}{\omega-\Omega_{\pm,k_1}}$ terms (as in Case~(II)), but $\bar{\mathbf{a}}^{(1)}$ include all the $\frac{1}{\omega-\Omega_{\pm,k}}$ terms with $k\neq k_1$ due to the non-diagonal matrix $\mathbf{X}$. 
We can understand this mode activation as being mediated by the dark mode via the multi-mode correlation $\xi_{k,k'}$. This interpretation is consistent with the effective inter-mode coupling that is dynamically facilitated by the dark modes, as shown in Eq.~\eqref{eq:integro-differential}.


\section{Multi-mode Activation in a 1D Cavity}\label{sec:1D_cavity}

In this section, we demonstrate the disorder-induced multimode activation using an effective one-dimensional Fabry–P\'erot (FP) cavity composed of two distributed Bragg reflector (DBR) mirrors that enclose a molecular layer and spacers (see Fig.~\ref{fig:fp_field}).

\subsection{The DBR mirrors and molecular layer}
We consider each DBR mirror to consist of 8 layers of alternating dielectric materials with distinct refractive indices. In our case, we use 4 layers of $\text{Ge}$ with thickness $d_\text{Ge}=0.42$ $\mu m$ and static dielectric constant $\epsilon_\text{Ge}=16.0$, and 4 layers of $\text{ZnO}$ with thickness $d_\text{ZnO}=0.34$ $\mu m$ and static dielectric constant $\epsilon_\text{ZnO}=3.24$.\cite{yang2023enabling} 
For simplicity, we assume the dielectric medium outside the cavity and the spacer inside the cavity have identical dielectric constant $\epsilon_0=1.00$. 
The cavity length enclosed by the DBR mirrors is set to be $L=24.0$ $\mu \text{m}$ and the cavity cross section area is $A=10^6$ $\mu\text{m}^2$. 

\begin{figure}[htbp]
    \centering
	\includegraphics[width=0.5\linewidth]{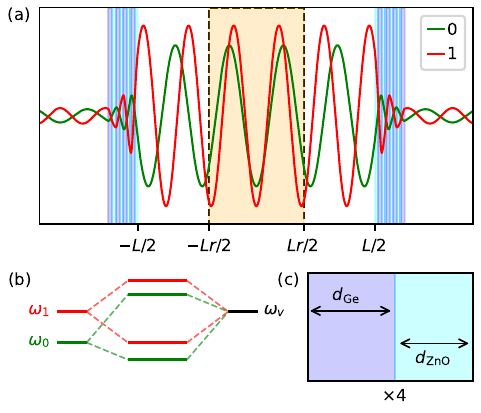}
    \caption{
    (a) Schematic diagram of an effective one-dimensional FP cavity of total length $L = 24.0\ \mu\text{m}$, formed by two DBR mirrors with thickness $d_\text{tot}=4\times 0.76\ \mu\text{m}$. The spatial profiles of the electric field for modes $\omega_0 = 1825\ \text{cm}^{-1}$ and $\omega_1 = 2184\ \text{cm}^{-1}$ are shown in green and red solid lines, respectively. A molecular layer is embedded in the central region $[-Lr/2, Lr/2]$ (shaded orange area), where $r$ denotes the ratio of the molecular layer thickness to the cavity length $L$. (b) Energy-level diagram illustrating the hybridization of the cavity photon modes $\omega_0$ and $\omega_1$ with the molecular excitation of frequency $\omega_v$. Each cavity photon mode couples to the molecules, leading to the hybridized states (center of the diagram). (c) Detailed view of a pair of dielectric layers within the DBR mirrors, consisting of Ge ($d_{\text{Ge}}$) and ZnO ($d_{\text{ZnO}}$) layers. Each DBR mirror consists of four pairs of such layers stacked together.}
    
    \label{fig:fp_field}
\end{figure}

To determine the photon modes of an empty FB cavity, we simulate the cavity ring-down spectrum (CRDS) using classical electrodynamics to extract the cavity photon modes in the absence of the molecular layer (see Appendix~\ref{appendix:CRDS}).\cite{taflove_computational_2005,zalicki_cavity_1995}
For the FP cavity specifically, we choose two sequential cavity photon modes with frequencies $\omega_0=1825\ \text{cm}^{-1}$ and $\omega_1=2184\ \text{cm}^{-1}$ and the corresponding linewidth $\gamma_0=0.53 \ \text{cm}^{-1}$ and $\gamma_1=0.60 \ \text{cm}^{-1}$. 
From the mode-specific electric field as obtained by the CRDS, we can also extract the spatial distribution ($\boldsymbol{\mathcal{G}}_0(x)$ and $\boldsymbol{\mathcal{G}}_1(x)$), respectively (shown in Fig.~\ref{fig:fp_field}).

Regarding the molecular layer, we define the central region inside the FP cavity as $\mathcal{V}_\mathrm{mol}=[-\frac{L}{2}{r},\frac{L}{2}{r}]$, which is sandwiched between two spacer layers. Here, the molecular layer ratio $r$ specifies the ratio of the total spacer thickness to the cavity length: $r=0$ corresponds to a cavity without molecules (an empty cavity), while $r=1$ corresponds to a cavity fully occupied by the molecular layer.  
The molecular number density within the molecular layer is set to be fixed as 
$\rho_\mathrm{mol}=7.06\times10^{9}~\mu\text{m}^{-3}$ ($\approx 11.7$ M), which is representative of pure liquid media.\cite{vergauwe2019modification}
We limit the molecular emitters to the single-excitation manifold and the molecular transition frequencies are set to be nearly resonant with mode 1 ($\omega_v = \omega_1$), and share a common lifetime, for which we set $\gamma_v = \gamma_1$.
The transition dipole moment can be written as $\boldsymbol{\mu}_j=\mu_0\mathbf{e}_j$ where $\mu_0 = 1.00~\text{Debye}$ and $\mathbf{e}_j$ is a randomly oriented unit vector that does not correlate with the positions. Due to random orientation, we have $\ov{u}_k=0$ and focus on disorder-induced phenomena in the following.



\subsection{Evaluation of Coupling Variance and Cross-Mode Correlations}
Having the mode-specific electric fields $\boldsymbol{\mathcal{G}}_k(x)$, we are now ready to determine the key parameters for the EOMs in the collective mode representation. 
The mode-specific coupling fluctuation $\sigma_k^2=\frac{1}{N}\sum_{j=1}^N |\boldsymbol{\mu}_j  \cdot  \boldsymbol{\mathcal{G}}_k(x_j)|^2$ can be determined by sampling the emitter positions $x_j$ within the molecular region $\mathcal{V}_\mathrm{mol}$.
In the large $N$ limit, we can evaluate the variance of the coupling variance for mode $k$ by
\begin{equation}
\sigma_k^2=\frac{\mu_0^2}{3N}\sum_{j=1}^N |\boldsymbol{\mathcal{G}}_k(x_j)|^2
\end{equation}
where the $1/3$ factor is from averaging over random orientation at each sampled $x_j$.
The cross-mode correlation can be evaluated by
\begin{equation}\label{eq:xi_discrete}
\xi_{k_1,k_2}=\frac{\sum_j \boldsymbol{\mathcal{G}}_{k_1}^*(x_j)\boldsymbol{\mathcal{G}}_{k_2}(x_j)}{\sqrt{\sum_j |\boldsymbol{\mathcal{G}}_{k_1}(x_j)|^2\sum_j |\boldsymbol{\mathcal{G}}_{k_2}(x_j)|^2}}
\end{equation}
which depends exclusively on the spatial distribution of the electric field. 
Note that, for simplicity, we assume the extracted cavity properties ($\omega_k$, $\gamma_k$, and $\mathcal{G}_k$) remain unchanged in the presence of the molecular layer and the molecular emission can be neglected.

In the continuum limit, we can further simplify the coupling variance and cross-mode correlation. Given that the molecular ensemble follows a constant density distribution within the molecular layer, we can evaluate the variance of coupling fluctuation for mode $k$ by
\begin{equation}\label{Eq:SIGMA_eval}
\sigma^2_k(r) = \frac{|\mu_0|^2}{3Lr}\int_{-Lr/2}^{Lr/2}|\boldsymbol{\mathcal{G}}_k(x)|^2dx
\end{equation}
The cross-mode correlation can therefore be evaluated by 
\begin{equation}\label{Eq:XI_eval}
\xi_{k_1,k_2}(r)=\frac{\int_{-Lr/2}^{Lr/2}\boldsymbol{\mathcal{G}}^*_{k_1}(x)\boldsymbol{\mathcal{G}}_{k_2}(x)dx}{\sqrt{\int_{-Lr/2}^{Lr/2}|\boldsymbol{\mathcal{G}}_{k_1}(x)|^2dx\int_{-Lr/2}^{Lr/2}|\boldsymbol{\mathcal{G}}_{k_2}(x)|^2dx}}
\end{equation}
Note that the coupling variance $\sigma_k(r)$ and the cross-mode correlation $\xi_{k_1,k_2}(r)$ are now explicit functions of the molecular layer ratio $r$ due to field inhomogeneity.

For a full FP cavity considered here ($r=1$) specifically, the coupling variance are $\sqrt{N}\sigma_0(r=1)=350~\text{cm}^{-1}$ and $\sqrt{N}\sigma_1(r=1)=370~\text{cm}^{-1}$, respectively. We also calculate the cross-mode correlation between the chosen modes {$\xi_{01}(r=1)=0.12$}.
Here we choose $\xi_{01}(r)$ to be a positive and real value, since we can absorb the complex phase of $\xi(r)$ into the external driving coupling $g_1$, which is irrelevant to the intensity of the cavity or dark modes.
In the following sections, $\xi(r)$ is used to denote $\xi_{01}(r)$ for brevity.

\subsection{Transient Dynamics}
\begin{figure}[htbp]
    \centering
    \includegraphics[width=0.5\linewidth]{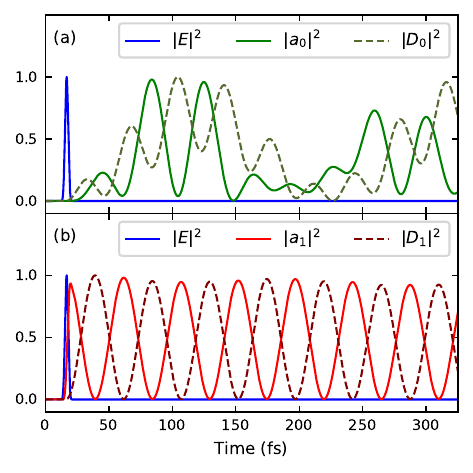}
    \caption{
    Transient dynamics of the cavity photon and its corresponding dark mode: (a) Mode 0: $|a_0|^2$, $|D_0|^2$, and (b) Mode 1: $|a_1|^2$, $|D_1|^2$ under a short pulse excitation (blue line). As the short pulse excites only mode 1, $a_1$ (red) and $D_1$ (dark red) exhibit oscillatory behavior (similar to Rabi oscillation) in (b). In (a), $D_0$ (green dashed line) is excited through disorder-induced coupling $\sqrt{N}\xi_{01}\sigma_{1}$, then exciting $a_0$ (green solid line). Note that the electric pulse peak intensity is normalized to unity, and the intensities for each mode $k$ ($|a_k|^2$ and $|D_k|^2$) are normalized by the maximum combined intensity of the cavity and dark modes, $\max(|a_k|^2, |D_k|^2)$.
    }\label{fig:dynamics}
\end{figure}

We first focus on the transient dynamics in response to a pulsed excitation.
The incident field takes the form of a Gaussian pulse: $\mathcal{E}_\mathrm{in}(t)=\cos(\omega_\mathrm{in}t)e^{-(t-t_0)^2/\tau^2}$ where  $\tau=3.34~\text{fs}$, $t_0=5\tau$, and $\omega_\mathrm{in}/(2\pi c)=2000~\text{cm}^{-1}$. Note that this short pulse maps to a broad Gaussian in the frequency domain, centered at $\omega_\mathrm{in}$ with $\omega_0<\omega_\mathrm{in}<\omega_1$, and therefore excites both cavity photon modes. 
To highlight energy transfer between modes, we instead constrain the input to drive only photon mode 1 by setting $g_0 = 0$ and $g_1 = 1$.

Figure~\ref{fig:dynamics} shows the transient dynamics of the expectation values $a_k(t)$ and $D_k(t)$ following the EOM in the collective mode representation.
In Fig.~\ref{fig:dynamics}(b), we observe that ${a}_1(t)$ is initially driven by the external field, followed by the oscillatory exchange between ${a}_1(t)$ and ${D}_1(t)$ modes through an effective coupling $\sqrt{N}\sigma_{1}$. 
This oscillatory exchange describes the mechanism behind disorder-induced spectral splitting. 
Fig.~\ref{fig:dynamics}(a) shows that ${a}_1(t)$ drives ${D}_0(t)$ through the term $\sqrt{N}\xi\sigma_{1}$, which only exists when $\xi\neq 0$. Sequentially, ${a}_0(t)$ is driven by ${D}_0(t)$ and results in nonzero ${a}_0(t)$. 
Note that the initial dynamics suggest that the excitation energy flow follows ${a}_1\rightarrow{D}_0\rightarrow{a}_0$.
We also notice that, since $g_0=0$, ${a}_0(t)$ is solely driven by the memory integration term in Eq.~\eqref{eq:integro-differential}.
This result confirms that dynamical coupling mediated by the dark mode leads to excitation energy transfer from the driven photon mode (mode 1) to other photon modes.

\subsection{Steady-state spectra and dark state population}

\begin{figure}[tbp]
    \centering
    \includegraphics[width=0.5\linewidth]{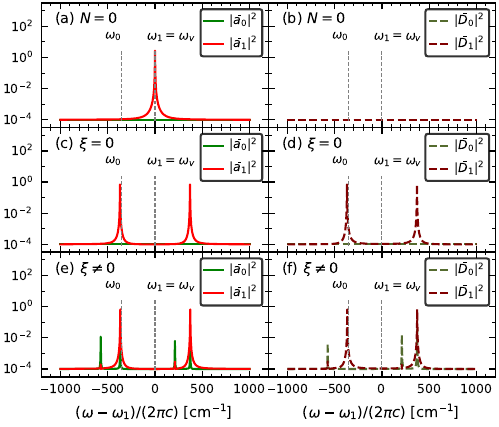}
    \caption{
    Steady-state spectra and dark state population on a logarithmic scale. Left column: (a, c, e) The intensity of cavity photon modes ($|\bar{a}_0|^2$, $|\bar{a}_1|^2$). Right column: (b, d, f) The population of dark states ($|\bar{D}_0|^2$, $|\bar{D}_1|^2$)
    The cavity modes ($\omega_0$ and $\omega_1$) are labeled by vertical dotted lines, and only mode 1 is driven by the external field. The molecule transition frequency is set to be $\omega_v=\omega_1$.
    For an empty cavity ($N=0$), we observe only one peak at $\omega_1$ in (a) and no dark mode population in (b). 
    For the $r=1$ case with $\xi =0$, we find that both $|\bar{a}_1|^2$ and $|\bar{D}_1|^2$ exhibit two peaks while $|\bar{a}_0|^2$ and $|\bar{D}_0|^2$ remain zero. The disorder-induced spectral splitting is determined by $\sqrt{N}\sigma_1=370 ~\text{cm}^{-1}$.
    For the $r=1$ case with $\xi =0.12$, we observed four peaks for both modes due to disorder-induced multimode activation, highlighting that nonzero $\xi$ leads to energy transfer from mode 1 to mode 0.
    }
    \label{fig:one_dimensional_all}
\end{figure}

\begin{figure}[htbp]
    \centering
    \includegraphics[width=0.5\linewidth]{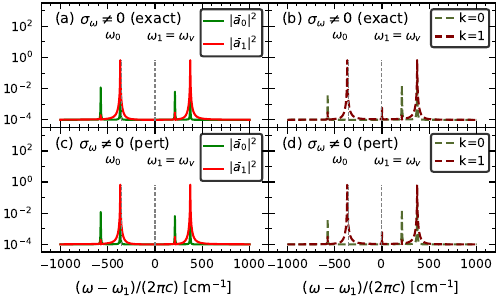}
    \caption{
    Comparison of steady-state spectra in the presence of frequency disorder obtained by (a,b) numerical solution and (c,d) perturbation. The cavity setting is the same as Fig.~\ref{fig:one_dimensional_all}. Left column: (a, c) The steady-state intensity of cavity photon modes ($|\bar{a}_0|^2$, $|\bar{a}_1|^2$) shows agreement with Fig.~\ref{fig:one_dimensional_all}(e). Right column: The dark state population can be captured by the second-order perturbations $\tilde{P}_k=|\bar{D}_k^{(0)} +\bar{D}_k^{(2)}|^2  + |\bar{G}_k^{(1)}|^2$ for both $k=0,1$. Note that the emergence of the dark state population at $\omega_v$ is solely contributed by the auxiliary mode $\bar{G}_k$ due to frequency disorder.}
    \label{fig:one_dimensional_all_2}
\end{figure}
We now turn our attention to the steady-state amplitude and the dark state population under continuous-wave excitation. 
As a reference, Fig.~\ref{fig:one_dimensional_all}(a)(b) shows an empty cavity ($r=0$) result. Since $\sigma_0(r=0)=\sigma_1(r=0)=0$ and $g_0=0$, we find that mode 0 ($|\ov{a}_0|^2$) is not excited, and both dark modes are not populated. $|\ov{a}_1|^2$ shows only one peak at $\omega_1=\omega_v$.

Figure~\ref{fig:one_dimensional_all}(c--f) exhibits the steady-state spectra for a full cavity ($r=1$) with and without the cross-mode correlation.
In Fig.~\ref{fig:one_dimensional_all}(c) and (d), we set $\xi=0$ and observe a two-peak structure for $|\ov{a}_1|^2$ and $|\ov{D}_1|^2$, attributed to disorder-induced spectral splitting. 
This two-peak structure for $|\ov{a}_1|^2$ and $|\ov{D}_1|^2$ at $\Omega_{\pm,1}=\omega_1\pm \sqrt{N}\sigma_1\approx2554, 1814~\text{cm}^{-1}$ corresponds to the Rabi oscillation in Fig.~\ref{fig:dynamics}(b). 
However, both the cavity photon and dark mode associated with mode 0 are not excited/populated because they are neither driven nor coupled to the other photon mode when there is no cross-mode correlation.

In Fig.~\ref{fig:one_dimensional_all}(e)(f), we use the cross-mode correlation $\xi=0.12$ as we evaluated using Eq.~\eqref{Eq:XI_eval} for $r=1$. For both the steady-state spectra and dark state populations, we observe four peaks located at $\Omega_{\pm,1}\approx\omega_1\pm \sqrt{N}\sigma_1$ and $\Omega_{\pm,0}\approx\frac{\omega_0+\omega_1}{2}\pm \sqrt{\frac{(\omega_0-\omega_1)^2}{4}+N\sigma_0^2}$. 
Note that, since the cross-mode correlation $\xi$ is small, the observed peak positions agree with the unperturbed results $\Omega_{\pm,0}\approx 2398, 1611~\text{cm}^{-1}$ based on Eq.~\eqref{eq:Omega_case2}. 
We emphasize that, for a small $\xi$, the peak positions do not depend on $\xi$, but the amplitudes of the disorder-induced peaks in $|\ov{a}_0|^2$ and $|\ov{D}_0|^2$ are determined by the magnitude of $\xi$.
This result confirms that, through disorder correlation ($\xi \neq 0$), photon excitation energy leaks from mode 1 (driven by the external field) into mode 0 (both $|\ov{a}_0|^2$ and $|\ov{D}_0|^2$) as induced by coupling disorder.

\subsection{Frequency disorder effect}
To examine the robustness of the observed spectrum under energetic disorder, we then introduce inhomogeneous broadening into the molecular ensemble.
In Fig.~\ref{fig:one_dimensional_all_2}, we model frequency disorder using the dual-frequency binomial distribution with $\sigma_\omega = 5\text{ cm}^{-1}$ (see Appendix~\ref{appendix:binomial_freq_disorder}). 
Comparing to Fig.~\ref{fig:one_dimensional_all}(e), we observe that the photon spectra $|\ov{a}_k|^2$ [Fig.~\ref{fig:one_dimensional_all_2}(a)] remain virtually unchanged.
However, we find that Fig.~\ref{fig:one_dimensional_all_2}(b) exhibits an additional peak at the molecular frequency $\omega_v$, suggesting that the frequency disorder redistributes spectral weight.

This additional peak can be captured using perturbation theory (assuming that $\sigma_\omega\ll\omega_1-\omega_0$ is relatively small). By treating the binomial frequency disorder as a perturbation, we find that the dark state population can be contributed by the collective mode corresponding to the frequency fluctuations. In Appendix~\ref{appendix:binomial_freq_disorder}, we approximate the dark state population as a combination of the dark modes $|\ov{D}_k^{(0)} +\ov{D}_k^{(2)}|^2$ and the auxiliary modes $|\ov{G}_k^{(1)}|^2$ (up to the first order). Here $\ov{G}_k$ denotes the steady-state amplitude of the auxiliary mode $\hat{G}_k\equiv\frac{1}{\sqrt{N}\sigma_k\sigma_\omega}\sum_j\delta\omega_j\delta u_{jk}^*\hat{c}_j$, which incorporates both the coupling and frequency disorder.
In  Fig.~\ref{fig:one_dimensional_all_2}(d), we show that the perturbation result overlays almost identically with the results in Fig.~\ref{fig:one_dimensional_all_2}(b) and the peak at $\omega_v$ is contributed by the $G_k$ mode. The perturbation result in Fig.~\ref{fig:one_dimensional_all_2}(c) exhibits agreement with the numerical result, suggesting that the DIMA spectrum is protected under small frequency disorder $\sigma_\omega$.

\section{Modulating Coupling Fluctuation and Disorder Correlation}\label{sec:trade-off}

In this section, we further explore how the steady-state spectra induced by DIMA can be modulated by changing the molecular--spacer ratio inside the cavity and the mirror properties comparing the DBR mirror and the ideal mirror modeled as a perfect electric conductor (PEC).
We again consider two cavity photon modes ($\omega_0$ and $\omega_1$), and only $\omega_1$ is driven by the incident field. 
To quantify the effective coupling between these photon modes, we recall the equation of motion for $a_0$ following Eq.~\eqref{eq:Dk_EOM_steady} and decompose the effective cross-mode coupling strength $\xi_{01}\sqrt{N}\sigma_{1}$ into the cross-mode correlation factor ($\xi=|\xi_{01}|$) and disorder-induced coupling factor ($\sqrt{N}\sigma_{1}$).

\subsection{Effective cross-mode coupling strength}
\begin{figure}[htbp]
    \centering
    \includegraphics[width=0.5\linewidth]{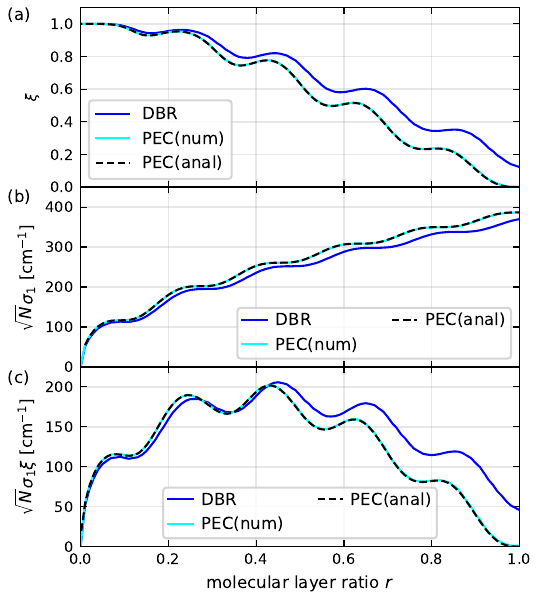}
    \caption{
    Comparison of the cross-mode correlations and effective disorder-induced coupling as a function of the molecular layer ratio $r$. 
    DBR(num) and  PEC(num) are numerical results obtained by FDTD simulation using the DBR mirrors and the ideal PEC mirror, respectively. PEC(anal) denotes the analytical results derived from Appendix~\ref{appendix:beating}.
    (a) The multi-mode correlation factor $\xi$ as a function of the molecular layer ratio $r$. The curves follow a general oscillatory decreasing trend. For $r\to0$ (empty cavity), Both mirror configurations converge to $\xi=1$. At $r\to1$, while $\xi\to0$ for the ideal mirror, $\xi\neq0$ for case A. 
    (b) The effective disorder-induced coupling $\sqrt{N}\sigma_1$ as a function of $r$. Both cases exhibit a similar increasing trend ($\propto \sqrt{r}$) with an oscillatory behavior.
    (c) The inter-mode coupling $\sqrt{N}\sigma_1\xi$ evaluated as the product of $\xi$ from (a) and $\sqrt{N}\sigma_1$ from (b). 
    We notice that the inter-mode coupling at $r = 1$ for case A does not drop to zero.
    This highlights an inherent trade-off between $\xi$ and $\sqrt{N}\sigma_k$ as a function of $r$, suggesting that the maximal disorder-induced activation occurs around $r \sim 0.45$.
    } 
    \label{fig:xi_all}
\end{figure}

Figure~\ref{fig:xi_all} exhibits the cross-mode correlation $|\xi|$ and the disorder-induced coupling $\sqrt{N}\sigma_1$ as a function of the molecular layer ratio ($r$), respectively.
We observe an overall pattern for both the ideal mirror and the DBR mirror that (a) $|\xi|$ decreases as $r$ grows, whereas (b) $\sqrt{N}\sigma_1$ increases with increasing $r$. 
As a result of a trade-off between these two factors, we find the maximal cross-mode coupling strength occurs around $r \sim 0.45$ (see Fig.~\ref{fig:xi_all}(c)).

\subsubsection{Cross-mode correlation factor}
To analyze how the cross-mode correlation changes as we change the molecular layer ratio ($r$), we first consider the limit $r\rightarrow0$ (almost empty cavity). In this limit, the molecular layer occupies only an infinitesimal region near the center ($x\rightarrow0$) and the mode-specific fields $\boldsymbol{\mathcal{G}}_{0}(x)$ and $\boldsymbol{\mathcal{G}}_{1}(x)$ are approximately independent of $x$ within this region (see Fig.~\ref{fig:fp_field}(a)). 
Thus, the cross-mode correlation approaches $\xi(r\rightarrow0)=1$ for both PEC and DBR mirrors.

In the limit $r=1$ (full cavity), we find that $|\xi|=0$ for the PEC mirror, whereas for the DBR mirror it approaches a nonzero finite value.
For the PEC mirror, the reflected field has a $\pi$-phase shift (a node at the PEC surface), suggesting that the electric fields are present only inside the cavity region and the mode-specific electric fields are orthogonal to each other within the cavity region (i.e., $\int_{-L/2}^{L/2}\boldsymbol{\mathcal{G}}^*_{0}(x)\boldsymbol{\mathcal{G}}_{1}(x)dx=0$). Therefore, the cross-mode correlation turns out to be zero for a full cavity formed by ideal mirrors. 

In contrast to an ideal PRC mirror, the phase shift of the reflected field for a DBR mirror is not equal to $\pi$. Consequently, the reflected and incident fields do not completely cancel at the mirror surface, which allows the field to penetrate into the DBR structure. 
We emphasize that, while the mode-specific electric fields are chosen to be orthogonal over the extended region including the cavity and the DBR mirrors, the cross-mode correlation is evaluated by the integration over the molecular layer region $[-Lr/2,Lr/2]$, which approaches a non-zero value in the limit $r=1$. 

\begin{figure*}[tbp]
    \centering
    \includegraphics[width=1.0\linewidth]{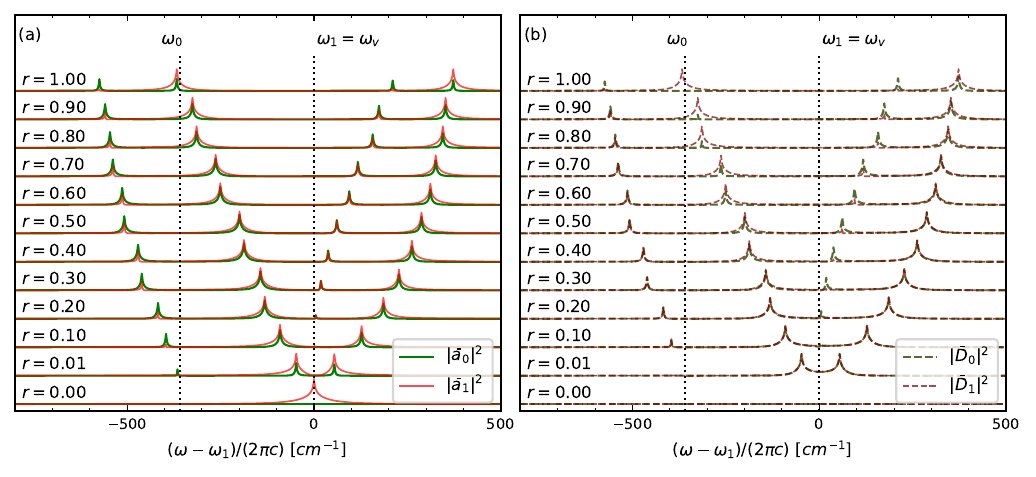}
    \caption{
    Steady-state spectra of the cavity photon modes (a) and the dark modes (b) on a logarithmic scale under various molecular layer ratios $r$. The cavity setup is the same as Fig.~\ref{fig:one_dimensional_all}. We scan from $r=1$ (top, full cavity) to $r\to0$ (bottom, almost empty cavity). (a) Decreasing main peak splitting in mode 1, which can be explained with decreasing $\sqrt{N_0r}\sigma_1(r)$. (b) Increasing overlap between $\bar{D}_0$ and $\bar{D}_1$ spectra, which can be explained with increasing $\xi(r)$. 
    These trends complement the trends of Fig.~\ref{fig:xi_all}(a)(b). }
    \label{fig:one_dimensional_11_cases}
\end{figure*}

For $0<r<1$, we find that $\xi$ exhibits an oscillatory behavior and, notably, the DBR mirror almost always enhances $\xi$.
The oscillatory behavior is attributed to the electric field inhomogeneity within the cavity region. Essentially, the overlap integration of two mode-specific fields of different frequencies $\omega_0\neq \omega_1$ leads to a beating frequency of $\frac{\omega_0+\omega_1}{2}$ (see Appendix~\ref{appendix:beating} for more details). Our analytical result for the PEC mirror agrees with numerical simulations.
More interestingly, the enhancement of the cross-mode correlation for the DBR mirror can be explained by the combined reduction of $\sigma_0$ and $\sigma_1$ in the denominator in Eq.~\eqref{Eq:XI_def}, which we discuss in the following.

\subsubsection{Disorder-induced coupling factor}
Figure~\ref{fig:xi_all}(b) exhibits that the disorder-induced coupling strength $\sqrt{N}\sigma_1$ tends to increase as $r$ grows. Here, the overall trend follows the $\sqrt{r}$ scaling. This trend is intuitive since the molecular layer has a constant density and the total number of emitters is proportional to the thickness of the molecular layer, i.e. $N\propto r$. On top of the increasing trend, the oscillatory behavior stems from the beating feature between two modes, following the functional form of $\sqrt{1+\mathrm{sinc}(kr)}$ (see Appendix~\ref{appendix:beating}). Overall, the disorder-induced coupling factor is dominated by the $\sqrt{N}$ scaling as the thickness grows.

Comparing the DBR and PEC mirrors, we find that the DBR mirror suppresses the disorder-induced coupling strength (in terms of $\sigma_k$). This is because the variance of coupling fluctuation is proportional to the electric field energy given by Eq.~\eqref{Eq:SIGMA_eval}. As the electric fields are extended into the DBR structure, the energy leak out of the molecular layer. Therefore, for a given $r$, $\sigma_k$ is overall reduced due to field penetration into the DBR mirror.


\subsubsection{Maximal effective coupling}
The effective coupling strength between mode 0 and 1 ($\sqrt{N}\sigma_1\xi$) is a combined result of Fig.~\ref{fig:xi_all}(a) and Fig.~\ref{fig:xi_all}(b). As illustrated in Fig.~\ref{fig:xi_all}(c), there is a fundamental trade-off between the cross-mode correlation $\xi$ and the number of molecules $\sqrt{N}$. When the molecular layer becomes thicker ($r$ increases), the disorder-induced coupling strength grows by $\sqrt{N}$, while the cross-mode correlation drops.
This trade-off relation points to a tunable knob for the DIMA effect: reducing the disorder-induced coupling strength ($\sqrt{N}\sigma_1$) can, in exchange, enhances the cross-mode correlation $\xi$. Consequently, the effective cross-mode coupling strength, quantified by $\sqrt{N}\sigma_1\xi$, reaches a maximum near $r \approx 0.45$.


\subsection{Steady-State Intensity Spectra}
With this trade-off relation in mind, we further explore how the steady-state spectrum varies as we change the molecular layer ratio $r$ within the DBR cavity. 
Figure~\ref{fig:one_dimensional_11_cases}(a) shows that the splitting between $\Omega_{\pm,k}$ peaks decreases as $r$ decreases, which mainly results from the reduction in the effective molecular number $\sqrt{N}\propto\sqrt{r}$. Specifically for mode 1, the splitting $\Omega_{+,1}-\Omega_{-,1}=2\sqrt{N}\sigma_1$ can be approximately captured by $\sqrt{r}$ when $r$ is large (see Appendix~\ref{appendix:beating}). 
We emphasize that, although the effective cross-mode coupling shows a non-monotonic behavior as $r$ is reduced, the separation between the spectral peaks is determined by the disorder-induced coupling strength ($\sqrt{N}\sigma_k$ in Eq.~\eqref{eq:Omega_case2}) and therefore exhibits the $\sqrt{r}$ scaling shown in Fig.~\ref{fig:xi_all})(b).

In Fig.~\ref{fig:one_dimensional_11_cases}(b), the dark states population shows a similar peak structure as (a) with the same position but different intensities. 
Interestingly, when $r$ decreases, $|\ov{D}_0|^2$ and $|\ov{D}_1|^2$ show a resemblance to each other. This resemblance is caused by the rise of $\xi$ as $r$ decreases. Physically, for a thin molecular layer, the electric field near $x=0$ is nearly a constant for both modes 0 and 1, so that $\xi\rightarrow1$. Thus, the dark modes $\hat{D}_0$ and $\hat{D}_1$ correspond to the same linear combination of the matter operators.
We also find that the intensity of the peaks at $\Omega_{\pm,0}$ tend to grow as $r$ decreases from $1$ to $0.5$, followed by a drop to $0$ as $r$ decreases from $0.5$ to $0.01$. This dark-state peak intensity corresponds to the turnover trend of the cross-mode coupling term $\sqrt{N}\sigma_1\xi$ in Fig.~\ref{fig:xi_all}(c).

\subsection{Integrated Intensity}
\begin{figure}[tbp]
\includegraphics[width=0.5\linewidth]{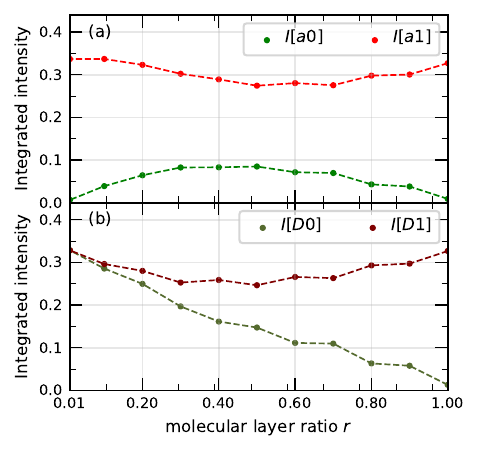}
    \caption{ 
    The integrated intensity under different molecular layer ratio $r$ values. We consider the two modes $\omega_0$ and $\omega_1$ but only mode 1 is active.     Molecules are on resonance with mode 1 $\omega_v=\omega_1$. 
    (a) {$I[a_k]$} are the photon mode intensities integrated over frequency $\int|\ov{a}_k(\omega)|^2d\omega$. We highlight that $I[a_0]$ increases then decreases as $r$ growes from 0 to 1, matching the trend of $\sqrt{N}\sigma_1\xi$ in Fig.~\ref{fig:xi_all}(c). 
    (b) {$I[D_k]$} are the dark state populations integrated over frequency $\int|\ov{D}_k(\omega)|^2d\omega$. Note that $I[D_0]$ converge to $I[D_1]$ at $r=0$ because $\xi(r=0)=1$, meaning that $D_0$ and $D_1$ modes are identical. }
    \label{fig:spectral_integral}
\end{figure}

To further quantify this manifestation of the trade-off relation, we calculate the integrated intensity by integrating the steady-state spectrum, $I[{a_k}]=\int d\omega |\bar{a}_k|^2$, and the dark state population, $I[{D_k}]=\int d\omega |\bar{D}_k|^2$, for various molecular layer ratios. In Fig.~\ref{fig:spectral_integral}, we observe that, at $r \approx 0.5$, $I[{a_0}]$ reaches a maximum and $I[a_1]$ shows a minimum, matching the turnover trend of the cross-mode coupling term $\sqrt{N}\sigma_1\xi$ as shown in Fig.~\ref{fig:xi_all}(c). 
This inverse correlation between $I[{a_0}]$ and $I[{a_1}]$ implies that, at the maximal effective coupling ($r\approx 0.45$), the DIMA effect facilitates photonic energy transfer from the excited photon mode 1 to the unexcited mode 0.

However, the dark modes ($I[{D_0}]$ and $I[{D_1}]$) do not show the same inverse behavior. 
While $I[{D_1}]$ exhibits a dip near $r \approx 0.5$, $I[{D_0}]$ decreases monotonically with $r$ and reaches the same value as $I_{D1}$ when $r\rightarrow0$. This discrepancy can be explained by the increasing correlation factor $\xi$ as shown in Fig.~\ref{fig:xi_all}(a). When $r\rightarrow0$ and $\xi\rightarrow1$, $\hat{D}_0$ and $\hat{D}_1$ become nearly degenerate and couple almost equally to the active photon mode 1. Therefore, the dark mode populations turn out to be equal as $r\rightarrow0$.

\section{Conclusion}\label{sec:conclusion}

To summarize this work, we have extended the theoretical framework of disorder-induced spectral splitting to a multi-mode Fabry-P\'erot cavity. By defining the collective dark modes and the cross-mode correlation, we demonstrate that excitation energy can be transferred from active photon modes (driven by an external field) to unexcited photon modes (decoupled from other photon modes) mediated by the enclosed molecular ensemble with orientation disorder. We exhibit a dynamical interplay between the cavity photon modes and their corresponding dark modes in terms of their transient behavior. We also confirm that the DIMA spectrum is robust under small frequency disorder, where local frequency disorder is effectively captured by an auxiliary matter mode, leaving the DIMA features intact.
Furthermore, we demonstrate the inverse relation between the disorder-induced coupling strength and the cross-mode correlation, leading to the maximal energy transfer when the cavity is around half-filled by the molecular layer.  
More importantly, our analysis of varying the molecular layer thickness and cavity mirrors offers a way to tune the DIMA effect and control the cross-mode photon energy transfer.

While this work considers an effective 1D Fabry-P\'erot cavity, the theory of DIMA can be applied to various platforms of strong light-matter coupling systems with disordered coupling. 
For example, a confocal cavity supports many nearly degenerate transverse modes that are decoupled due to their spatial symmetry.\cite{wickenbrock_collective_2013} The cross-mode coupling observed in atom cloud experiments can be explained and manipulated by the DIMA framework. In addition, for a plasmonic nanoparticle that supports multiple surface plasmonic modes,\cite{ghosh_energy_2025,bhandari_plasmon-exciton_2026} coating a molecular ensemble can mediate energy transfer among these modes. Further advancements in this direction facilitate controlled manipulation of these photon modes.



\begin{acknowledgments}
The authors acknowledge support from the University of Notre Dame and the Material Sciences and Engineering (MSE) fellowship. This material is based upon work supported by the National Science Foundation under Grant Number: CHE-2542091. We are grateful for discussions regarding the JPCL DISS framework.
\end{acknowledgments}
\bibliography{references}

@book{cohen-tannoudji_photons_1997,
    address = {New York},
    title = {Photons and {Atoms}: {Introduction} to {Quantum} {Electrodynamics}},
    isbn = {978-0-471-18433-1},
    shorttitle = {Photons and {Atoms}},
    url = {https://www.wiley.com/en-us/Photons+and+Atoms%3A+Introduction+to+Quantum+Electrodynamics-p-9780471184331},
    urldate = {2018-01-27},
    publisher = {Wiley},
    author = {Cohen-Tannoudji, Claude and Dupont-Roc, Jacques and Grynberg, Gilbert},
    year = {1997},
}

@book{taflove_computational_2005,
    address = {Boston},
    edition = {3rd ed. edition},
    title = {Computational {Electrodynamics}: {The} {Finite}-{Difference} {Time}-{Domain} {Method}, {Third} {Edition}},
    isbn = {978-1-58053-832-9},
    shorttitle = {Computational {Electrodynamics}},
    publisher = {Artech House Publishers},
    author = {Taflove, Allen and Hagness, Susan C.},
    month = jun,
    year = {2005},
}

@article{wickenbrock_collective_2013,
  title={Collective strong coupling in multimode cavity QED},
  author={Wickenbrock, Arne and Hemmerling, Michal and Robb, Gordon RM and Emary, Clive and Renzoni, Ferruccio},
  journal={Physical Review A—Atomic, Molecular, and Optical Physics},
  volume={87},
  number={4},
  pages={043817},
  year={2013},
  publisher={APS}
}

@article{huang_microcavity-assisted_2024,
    title = {Microcavity-assisted multi-resonant metasurfaces enabling versatile wavefront engineering},
    volume = {15},
    copyright = {2024 The Author(s)},
    issn = {2041-1723},
    url = {https://www.nature.com/articles/s41467-024-54057-9},
    doi = {10.1038/s41467-024-54057-9},
     
    number = {1},
    urldate = {2026-05-31},
    journal = {Nature Communications},
    publisher = {Nature Publishing Group},
    author = {Huang, Shih-Hsiu and Su, Hsiu-Ping and Chen, Chao-Yun and Lin, Yu-Chun and Yang, Zijin and Shi, Yuzhi and Song, Qinghua and Wu, Pin Chieh},
    month = nov,
    year = {2024},
    pages = {9658},
}

@article{reshef_multiresonant_2019,
    title = {Multiresonant {High}-{Q} {Plasmonic} {Metasurfaces}},
    volume = {19},
    issn = {1530-6984},
    url = {https://doi.org/10.1021/acs.nanolett.9b02638},
    doi = {10.1021/acs.nanolett.9b02638},
    number = {9},
    urldate = {2026-06-06},
    journal = {Nano Letters},
    publisher = {American Chemical Society},
    author = {Reshef, Orad and Saad-Bin-Alam, Md and Huttunen, Mikko J. and Carlow, Graham and Sullivan, Brian T. and Ménard, Jean-Michel and Dolgaleva, Ksenia and Boyd, Robert W.},
    month = sep,
    year = {2019},
    pages = {6429--6434},
}

@article{prado_atom-mediated_2011,
    title = {Atom-mediated effective interactions between modes of a bimodal cavity},
    volume = {84},
    copyright = {http://link.aps.org/licenses/aps-default-license},
    issn = {1050-2947, 1094-1622},
    url = {https://link.aps.org/doi/10.1103/PhysRevA.84.053839},
    doi = {10.1103/PhysRevA.84.053839},
     
    number = {5},
    urldate = {2026-07-08},
    journal = {Physical Review A},
    author = {Prado, F. O. and Luiz, F. S. and Villas-Bôas, J. M. and Alcalde, A. M. and Duzzioni, E. I. and Sanz, L.},
    month = nov,
    year = {2011},
    pages = {053839},
}

@article{ribeiro2018polariton,
  title={Polariton chemistry: controlling molecular dynamics with optical cavities},
  author={Ribeiro, Raphael F and Mart{\'\i}nez-Mart{\'\i}nez, Luis A and Du, Matthew and Campos-Gonzalez-Angulo, Jorge and Yuen-Zhou, Joel},
  journal={Chemical science},
  volume={9},
  number={30},
  pages={6325--6339},
  year={2018},
  publisher={The Royal Society of Chemistry}
}

@article{gonzalez2024light,
  title={Light--matter interactions in quantum nanophotonic devices},
  author={Gonz{\'a}lez-Tudela, Alejandro and Reiserer, Andreas and Garc{\'\i}a-Ripoll, Juan Jos{\'e} and Garc{\'\i}a-Vidal, Francisco J},
  journal={Nature Reviews Physics},
  volume={6},
  number={3},
  pages={166--179},
  year={2024},
  publisher={Nature Publishing Group UK London}
}

@article{engelhardt2023polariton,
  title={Polariton localization and dispersion properties of disordered quantum emitters in multimode microcavities},
  author={Engelhardt, Georg and Cao, Jianshu},
  journal={Physical review letters},
  volume={130},
  number={21},
  pages={213602},
  year={2023},
  publisher={APS}
}

@article{catuto2025interplay,
  title={Interplay between static and dynamic disorder: Contrasting effects on dark state population inside a cavity},
  author={Catuto, Robert F and Chen, Hsing-Ta},
  journal={The Journal of Chemical Physics},
  volume={162},
  number={20},
  year={2025},
  publisher={AIP Publishing}
}

@article{yang2023enabling,
  title={Enabling multiple intercavity polariton coherences by adding quantum confinement to cavity molecular polaritons},
  author={Yang, Zimo and Bhakta, Harsh H and Xiong, Wei},
  journal={Proceedings of the National Academy of Sciences},
  volume={120},
  number={1},
  pages={e2206062120},
  year={2023},
  publisher={National Academy of Sciences}
}

@article{mandal2023microscopic,
  title={Microscopic theory of multimode polariton dispersion in multilayered materials},
  author={Mandal, Arkajit and Xu, Ding and Mahajan, Ankit and Lee, Joonho and Delor, Milan and Reichman, David R},
  journal={Nano Letters},
  volume={23},
  number={9},
  pages={4082--4089},
  year={2023},
  publisher={ACS Publications}
}

@article{sun2025exploring,
  title={Exploring the delocalization of dark states in a multimode optical cavity},
  author={Sun, Kunyang and Du, Matthew and Yuen-Zhou, Joel},
  journal={The Journal of Physical Chemistry C},
  volume={129},
  number={21},
  pages={9837--9843},
  year={2025},
  publisher={ACS Publications}
}

@article{georgiou2021observation,
  title={Observation of photon-mode decoupling in a strongly coupled multimode microcavity},
  author={Georgiou, Kyriacos and McGhee, Kirsty E and Jayaprakash, Rahul and Lidzey, David G},
  journal={The Journal of Chemical Physics},
  volume={154},
  number={12},
  year={2021},
  publisher={AIP Publishing}
}

@article{tavis_exact_1968,
    title = {Exact {Solution} for an {N} -{Molecule}—{Radiation}-{Field} {Hamiltonian}},
    volume = {170},
    copyright = {http://link.aps.org/licenses/aps-default-license},
    issn = {0031-899X},
    url = {https://link.aps.org/doi/10.1103/PhysRev.170.379},
    doi = {10.1103/PhysRev.170.379},
     
    number = {2},
    urldate = {2024-06-18},
    journal = {Physical Review},
    author = {Tavis, Michael and Cummings, Frederick W.},
    month = jun,
    year = {1968},
    pages = {379--384},
}

@article{houdre_vacuum-field_1996,
    title = {Vacuum-field {Rabi} splitting in the presence of inhomogeneous broadening: {Resolution} of a homogeneous linewidth in an inhomogeneously broadened system},
    volume = {53},
    issn = {1050-2947, 1094-1622},
    shorttitle = {Vacuum-field {Rabi} splitting in the presence of inhomogeneous broadening},
    url = {https://link.aps.org/doi/10.1103/PhysRevA.53.2711},
    doi = {10.1103/PhysRevA.53.2711},
     
    number = {4},
    urldate = {2020-06-03},
    journal = {Physical Review A},
    author = {Houdré, R. and Stanley, R. P. and Ilegems, M.},
    month = apr,
    year = {1996},
    pages = {2711--2715},
}

@misc{zhou_interplay_2022,
    title = {The {Interplay} between {Disorder}, {Local} {Relaxation} and {Collective} {Behaviors} for an ensemble of emitters outside vs inside cavity},
    url = {http://arxiv.org/abs/2211.16325},
    urldate = {2023-02-07},
    publisher = {arXiv},
    author = {Zhou, Zeyu and Chen, Hsing-Ta and Subotnik, Joseph E. and Nitzan, Abraham},
    month = nov,
    year = {2022},
    note = {arXiv:2211.16325 [physics]},
}

@misc{ying_microscopic_2023,
    title = {Microscopic {Theory} of {Vibrational} {Polariton} {Chemistry}},
    url = {http://arxiv.org/abs/2305.05005},
    urldate = {2023-10-28},
    publisher = {arXiv},
    author = {Ying, Wenxiang and Taylor, Michael A. D. and Huo, Pengfei},
    month = may,
    year = {2023},
    note = {arXiv:2305.05005 [physics, physics:quant-ph]},
}

@article{ying_resonance_2024,
    title = {Resonance theory of vibrational polariton chemistry at the normal incidence},
    copyright = {De Gruyter expressly reserves the right to use all content for commercial text and data mining within the meaning of Section 44b of the German Copyright Act.},
    issn = {2192-8614},
    url = {https://www.degruyter.com/document/doi/10.1515/nanoph-2023-0685/html},
    doi = {10.1515/nanoph-2023-0685},
     
    urldate = {2024-05-04},
    journal = {Nanophotonics},
    publisher = {De Gruyter},
    author = {Ying, Wenxiang and Taylor, Michael A. D. and Huo, Pengfei},
    month = feb,
    year = {2024},
}

@article{ying_microscopic_2025,
    title = {Microscopic theory of polariton group velocity renormalization},
    volume = {16},
    copyright = {2025 The Author(s)},
    issn = {2041-1723},
    url = {https://www.nature.com/articles/s41467-025-62276-x},
    doi = {10.1038/s41467-025-62276-x},
     
    number = {1},
    urldate = {2025-09-09},
    journal = {Nature Communications},
    publisher = {Nature Publishing Group},
    author = {Ying, Wenxiang and Chng, Benjamin X. K. and Delor, Milan and Huo, Pengfei},
    month = jul,
    year = {2025},
    pages = {6950},
}

@article{wright_versatile_2023,
    title = {A versatile platform for gas-phase molecular polaritonics},
    volume = {159},
    issn = {0021-9606},
    url = {https://doi.org/10.1063/5.0170326},
    doi = {10.1063/5.0170326},
    number = {16},
    urldate = {2024-04-18},
    journal = {The Journal of Chemical Physics},
    author = {Wright, Adam D. and Nelson, Jane C. and Weichman, Marissa L.},
    month = oct,
    year = {2023},
    pages = {164202},
}

@article{balasubrahmaniyam_coupling_2021,
    title = {Coupling and decoupling of polaritonic states in multimode cavities},
    volume = {103},
    issn = {2469-9950, 2469-9969},
    url = {https://link.aps.org/doi/10.1103/PhysRevB.103.L241407},
    doi = {10.1103/PhysRevB.103.L241407},
     
    number = {24},
    urldate = {2026-06-12},
    journal = {Physical Review B},
    author = {Balasubrahmaniyam, M. and Genet, Cyriaque and Schwartz, Tal},
    month = jun,
    year = {2021},
    pages = {L241407},
}

@article{crookes_collective_2025,
    title = {Collective multimode strong coupling in plasmonic nanocavities},
    volume = {14},
    copyright = {© 2025 The Authors},
    issn = {2192-8614},
    url = {https://onlinelibrary.wiley.com/doi/abs/10.1515/nanoph-2024-0618},
    doi = {10.1515/nanoph-2024-0618},
     
    number = {11},
    urldate = {2026-06-12},
    journal = {Nanophotonics},
    author = {Crookes, Angus and Yuen, Ben and Demetriadou, Angela},
    year = {2025},
    note = {\_eprint: https://onlinelibrary.wiley.com/doi/pdf/10.1515/nanoph-2024-0618},
    pages = {2065--2073},
}

@article{rahmanian_koshkaki_exciton-polariton_2026,
    title = {Exciton-polariton dynamics in multilayered materials},
    volume = {17},
    copyright = {2026 The Author(s)},
    issn = {2041-1723},
    url = {https://www.nature.com/articles/s41467-026-72699-9},
    doi = {10.1038/s41467-026-72699-9},
     
    number = {1},
    urldate = {2026-08-30},
    journal = {Nature Communications},
    publisher = {Nature Publishing Group},
    author = {Rahmanian Koshkaki, Saeed and Manjalingal, Arshath and Blackham, Logan and Mandal, Arkajit},
    month = may,
    year = {2026},
    pages = {6156},
}

@article{ribeiro_multimode_2022,
    title = {Multimode polariton effects on molecular energy transport and spectral fluctuations},
    volume = {5},
    issn = {2399-3669},
    url = {https://www.nature.com/articles/s42004-022-00660-0},
    doi = {10.1038/s42004-022-00660-0},
     
    number = {1},
    urldate = {2023-03-22},
    journal = {Communications Chemistry},
    author = {Ribeiro, Raphael F.},
    month = apr,
    year = {2022},
    pages = {48},
}

@article{morton_theoretical_2011,
    title = {Theoretical {Studies} of {Plasmonics} using {Electronic} {Structure} {Methods}},
    volume = {111},
    issn = {0009-2665, 1520-6890},
    url = {https://pubs.acs.org/doi/10.1021/cr100265f},
    doi = {10.1021/cr100265f},
     
    number = {6},
    urldate = {2019-06-15},
    journal = {Chemical Reviews},
    author = {Morton, Seth M. and Silverstein, Daniel W. and Jensen, Lasse},
    month = jun,
    year = {2011},
    pages = {3962--3994},
}

@article{xiang_intermolecular_2020,
    chapter = {Report},
    title = {Intermolecular vibrational energy transfer enabled by microcavity strong light–matter coupling},
    volume = {368},
    copyright = {Copyright © 2020 The Authors, some rights reserved; exclusive licensee American Association for the Advancement of Science. No claim to original U.S. Government Works. http://www.sciencemag.org.proxy.library.upenn.edu/about/science-licenses-journal-article-reuseThis is an article distributed under the terms of the Science Journals Default License.},
    issn = {0036-8075, 1095-9203},
    url = {https://science-sciencemag-org.proxy.library.upenn.edu/content/368/6491/665},
    doi = {10.1126/science.aba3544},
     
    number = {6491},
    urldate = {2020-08-08},
    journal = {Science},
    publisher = {American Association for the Advancement of Science},
    author = {Xiang, Bo and Ribeiro, Raphael F. and Du, Matthew and Chen, Liying and Yang, Zimo and Wang, Jiaxi and Yuen-Zhou, Joel and Xiong, Wei},
    month = may,
    year = {2020},
    pages = {665--667},
}

@article{li_disorder-induced_2025,
    title = {Disorder-{Induced} {Spectral} {Splitting} versus {Rabi} {Splitting} under {Strong} {Light}-{Matter} {Coupling}},
    url = {https://pubs.acs.org/doi/abs/10.1021/acs.jpclett.5c01302},
    doi = {10.1021/acs.jpclett.5c01302},
    urldate = {2025-06-24},
    journal = {The Journal of Physical Chemistry Letters},
    publisher = {American Chemical Society},
    author = {Li, Wei-Kuo and Chen, Hsing-Ta},
    month = jun,
    year = {2025},
    pages = {6728--6733},
}

@article{vergauwe2019modification,
  title={Modification of enzyme activity by vibrational strong coupling of water},
  author={Vergauwe, Robrecht MA and Thomas, Anoop and Nagarajan, Kalaivanan and Shalabney, Atef and George, Jino and Chervy, Thibault and Seidel, Marcus and Devaux, Elo{\"\i}se and Torbeev, Vladimir and Ebbesen, Thomas W},
  journal={Angewandte Chemie International Edition},
  volume={58},
  number={43},
  pages={15324--15328},
  year={2019},
  publisher={Wiley Online Library}
}

@article{rokaj_lightmatter_2018,
    title = {Light–matter interaction in the long-wavelength limit: no ground-state without dipole self-energy},
    volume = {51},
    issn = {0953-4075, 1361-6455},
    shorttitle = {Light–matter interaction in the long-wavelength limit},
    url = {https://iopscience.iop.org/article/10.1088/1361-6455/aa9c99},
    doi = {10.1088/1361-6455/aa9c99},
     
    number = {3},
    urldate = {2024-08-02},
    journal = {Journal of Physics B: Atomic, Molecular and Optical Physics},
    author = {Rokaj, Vasil and Welakuh, Davis M and Ruggenthaler, Michael and Rubio, Angel},
    month = feb,
    year = {2018},
    pages = {034005},
}

@article{zalicki_cavity_1995,
    title = {Cavity ring-down spectroscopy for quantitative absorption measurements},
    volume = {102},
    issn = {0021-9606, 1089-7690},
    url = {https://pubs.aip.org/jcp/article/102/7/2708/481691/Cavity-ring-down-spectroscopy-for-quantitative},
    doi = {10.1063/1.468647},
     
    number = {7},
    urldate = {2026-09-18},
    journal = {The Journal of Chemical Physics},
    author = {Zalicki, Piotr and Zare, Richard N.},
    month = feb,
    year = {1995},
    pages = {2708--2717},
}

@article{ghosh_energy_2025,
    title = {Energy transfer and radiation damping in gold–{MAPbI3} heterostructures},
    volume = {16},
    issn = {2041-6520},
    url = {https://doi.org/10.1039/d5sc05386b},
    doi = {10.1039/d5sc05386b},
    number = {48},
    urldate = {2026-10-01},
    journal = {Chemical Science},
    author = {Ghosh, Bikram and Shingote, Ajinkya and Bhandari, Janak and Hartland, Gregory V.},
    month = dec,
    year = {2025},
    pages = {23012--23018},
}

@article{bhandari_plasmon-exciton_2026,
    title = {Plasmon-{Exciton} {Coupling} and {Dephasing} in {Hybrid} {Au} {Nanostructure}/{J}-{Aggregate} {Systems}},
    volume = {130},
    copyright = {https://creativecommons.org/licenses/by-nc-nd/4.0/},
    issn = {1932-7447, 1932-7455},
    url = {https://pubs.acs.org/doi/10.1021/acs.jpcc.6c02759},
    doi = {10.1021/acs.jpcc.6c02759},
     
    number = {29},
    urldate = {2026-07-24},
    journal = {The Journal of Physical Chemistry C},
    author = {Bhandari, Janak and Catuto, Robert and Zhang, Zhumin and Smith, Bradley D. and Chen, Hsing-Ta and Hartland, Gregory V.},
    month = jul,
    year = {2026},
    pages = {10287--10294},
}

@article{kristensen_modes_2014,
    title = {Modes and {Mode} {Volumes} of {Leaky} {Optical} {Cavities} and {Plasmonic} {Nanoresonators}},
    volume = {1},
    issn = {2330-4022, 2330-4022},
    url = {https://pubs.acs.org/doi/10.1021/ph400114e},
    doi = {10.1021/ph400114e},
     
    number = {1},
    urldate = {2020-01-05},
    journal = {ACS Photonics},
    author = {Kristensen, Philip Trøst and Hughes, Stephen},
    month = jan,
    year = {2014},
    pages = {2--10},
}

@article{xu_ultrafast_2023,
    title = {Ultrafast imaging of polariton propagation and interactions},
    volume = {14},
    copyright = {2023 The Author(s)},
    issn = {2041-1723},
    url = {https://www.nature.com/articles/s41467-023-39550-x},
    doi = {10.1038/s41467-023-39550-x},
     
    number = {1},
    urldate = {2024-05-03},
    journal = {Nature Communications},
    publisher = {Nature Publishing Group},
    author = {Xu, Ding and Mandal, Arkajit and Baxter, James M. and Cheng, Shan-Wen and Lee, Inki and Su, Haowen and Liu, Song and Reichman, David R. and Delor, Milan},
    month = jun,
    year = {2023},
    pages = {3881},
}

@article{xiang_molecular_2024,
    title = {Molecular {Polaritons} for {Chemistry}, {Photonics} and {Quantum} {Technologies}},
    volume = {124},
    issn = {0009-2665},
    url = {https://doi.org/10.1021/acs.chemrev.3c00662},
    doi = {10.1021/acs.chemrev.3c00662},
    number = {5},
    urldate = {2024-03-28},
    journal = {Chemical Reviews},
    publisher = {American Chemical Society},
    author = {Xiang, Bo and Xiong, Wei},
    month = mar,
    year = {2024},
    pages = {2512--2552},
}

@article{ebbesen_hybrid_2016,
    title = {Hybrid {Light}–{Matter} {States} in a {Molecular} and {Material} {Science} {Perspective}},
    volume = {49},
    issn = {0001-4842},
    url = {https://doi.org/10.1021/acs.accounts.6b00295},
    doi = {10.1021/acs.accounts.6b00295},
    number = {11},
    urldate = {2026-10-01},
    journal = {Accounts of Chemical Research},
    author = {Ebbesen, Thomas W.},
    month = nov,
    year = {2016},
    pages = {2403--2412},
}

@article{xiong_molecular_2023,
    title = {Molecular {Vibrational} {Polariton} {Dynamics}: {What} {Can} {Polaritons} {Do}?},
    issn = {0001-4842, 1520-4898},
    shorttitle = {Molecular {Vibrational} {Polariton} {Dynamics}},
    url = {https://pubs.acs.org/doi/10.1021/acs.accounts.2c00796},
    doi = {10.1021/acs.accounts.2c00796},
     
    urldate = {2023-03-30},
    journal = {Accounts of Chemical Research},
    author = {Xiong, Wei},
    month = mar,
    year = {2023},
    pages = {acs.accounts.2c00796},
}

@article{fregoni_theoretical_2022,
    title = {Theoretical {Challenges} in {Polaritonic} {Chemistry}},
    volume = {9},
    url = {https://doi.org/10.1021/acsphotonics.1c01749},
    doi = {10.1021/acsphotonics.1c01749},
    number = {4},
    urldate = {2024-08-01},
    journal = {ACS Photonics},
    publisher = {American Chemical Society},
    author = {Fregoni, Jacopo and Garcia-Vidal, Francisco J. and Feist, Johannes},
    month = apr,
    year = {2022},
    pages = {1096--1107},
}
\appendix

\section{Cavity Ring-Down Spectrum}\label{appendix:CRDS}
To characterize the photon mode properties of the DBR cavity, we simulate the cavity ring-down spectrum (CRDS) using classical electrodynamics, specifically the finite-difference time-domain (FDTD) method.
For the FDTD simulation, we assume the dielectric constant is $\epsilon_{\text{sol}}=1.00$ within an empty cavity and the electromagnetic field is not influenced by the molecular layer, so that the extracted cavity properties ($\omega_k$, $\Gamma_k$, and $\mathcal{G}_k$) remain independent of the molecular layer.

We place a light source of a Gaussian pulse $
E_y(t) \propto\exp[-(t-t_0)^2 / \tau^2]\cos(\omega_p t)
$
at the center of the cavity ($x=0$), where $\omega_p/(2\pi c)=2000$ cm$^{-1}$, $\tau^{-1}/c= 707$ cm$^{-1}$ and $t_0 = 5\sqrt{2}\tau/2$.
To detect the electromagnetic field, we place $n_{\mathrm{monitors}}=3$ electric field monitors ($i=0,1,2$) at
$x_i/L=-0.35,\ 0.0,\ 0.15$ within the cavity (where nodes are avoided).
The electric field obtained from the simulation is $\mathbf{E}_\text{num}(x_i,t)$. We then choose a $t_\text{cutoff}$ and only take $\mathbf{E}_\text{num}(x_i,t \geq t_\text{cutoff})$  as the ring-down signal. The cutoff time $t_\text{cutoff}$ is chosen to be 3.753 ps, which is large enough that the light source no longer pumps energy to the system and small enough such that the ring-down signal has not yet decayed. For convenience, we introduce an artificial damping constant $\eta^{-1}=25.02$ ps that is used to further damp the signal.
We then perform a Fourier transform to the angular-frequency domain,
\begin{equation}
\mathbf{E}_{\text{num}}(x_i, \omega)
= \int_{t_\mathrm{cutoff}}dt \mathbf{E}_{\text{num}}(x_i, t) e^{i\omega t-\eta t},
\end{equation}
and then define the spectrum as the squared magnitude $I(x_i,\omega)\equiv\left|{\mathbf{E}}_{\text{num}}(x_i, \omega)\right|^2$

We fit the spectra $I(\mathbf{r}_i,\omega)$ to a series of Lorentzians to extract mode angular frequencies $\omega_k$ and linewidth $\gamma_k \equiv \Gamma_k/\hbar$,
\begin{equation}
I(x_i,\omega) = \sum_k \frac{A(x_i)}{(\omega-\omega_k)^2 + \gamma_k^2}
\label{Eq:fitting}
\end{equation}
Although the fitted amplitude $A(x_i)$ of each monitor depends on the position, the angular frequencies of their extracted modes $\omega_k$ and their linewidth $\gamma_k$ do not have spatial dependence. 
Note that the fitted linewidth $\gamma_k$ from the ring-down signal is overestimated due to the artificial damping 
$\eta$ introduced in the time-domain signals. 
We therefore correct the fitted linewidth of mode $k$ by 
$\gamma_k \rightarrow \gamma_k - \eta$.
We tested multiple monitors to ensure that no modes are accidentally missed (if $x_i$ happens to be at the nodes of some mode). 

\begin{figure}[tbp]
    \centering
    \includegraphics[width=0.5\linewidth]{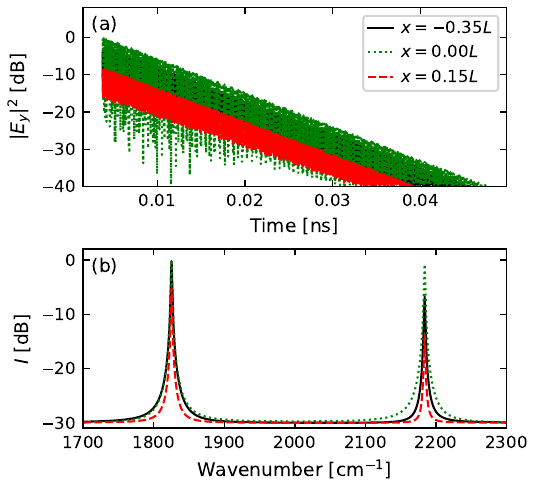}
    \caption{
    Simulated ring-down of the DBR cavity. (a) Decay of the field energy $|E_y|^2$ at three monitor positions($x/L=$-0.35, 0.00, 0.15), shown as the local maxima of the ring-down trace, normalized to the global maximum and expressed in dB; a $-40$ dB floor is applied, below which the field has decayed into the numerical noise. (b) Calculated spectra $I(x_i,\omega)$ at the same positions of the ring-down field in (a), normalized to the global maximum and plotted in dB with a $-30$ dB floor added so the spectrum remains well defined near zero.
    }
    \label{fig:fp_spectra}
\end{figure}

For the DBR cavity we considered, we extracted two sequential cavity photon modes: $\omega_0=1825\ \text{cm}^{-1}$ with $\gamma_0=0.53 \ \text{cm}^{-1}$ and $\omega_1=2184\ \text{cm}^{-1}$ with $\gamma_1=0.60 \ \text{cm}^{-1}$. 

From the electric field obtained via CRDS, we can now extract the mode-specific electric field by 
\begin{equation}
    \mathcal{E}_k(x) \equiv  \int_{t_\mathrm{cutoff}} dt \mathbf{E}_\text{num}(x,t) e^{i\omega_kt-\eta t}.
\end{equation}
To obtain the spatial dependence factor, $\mathcal{G}_k(x)=\mathcal{E}_{k}(x)/a^\circ_k$, we need to estimate $a^\circ_k=\langle\hat{a}^\circ_k\rangle$ (the classical correspondence of the photon operator in an empty cavity).
We equate the CQED expression of the internal energy of the $\omega_k$ mode $U_k=\hbar\omega_k(|a^\circ_k|^2+\frac{1}{2})$ to the classical electromagnetic energy enclosed in the cavity $U_k=2\int_\mathcal{V} \epsilon(x) |\mathcal{E}_k(x)|^2 dx \times S$ .  Note that here $\mathcal{V}=[-\frac{L+d_m}{2},\frac{L+d_m}{2}]$ denotes the total region of the cavity and the mirrors and $S$ is the mirror cross-section area. Unlike a perfect mirror, the electromagnetic fields within the DBR mirror are not zero, so we should include the mirror region when we calculate the energy. Thus, by numerically calculating the electromagnetic energy within the $\omega_k$ mode, one can calculate $|a^\circ_k|$. Then we divide $\mathcal{E}_{k}(x)$ by $|a^\circ_k|$ to get $\mathcal{G}_k(x)$ up to a phase.

\section{Analytical expression of the cross-mode coupling for an ideal mirror cavity}\label{appendix:beating}
For an ideal mirror cavity of length $L=24$  $\mu m$ and area $S=10^6$ $\mu m^2$  filled with a dielectric of $\epsilon_r=1.00$, the mode-specific electric fields take the form $\boldsymbol{\mathcal{G}}(x) = A_n\cos(n\pi x/L)\boldsymbol{y}$ where $n=1,3,5,7,\cdots$ for the odd modes in a symmetric cavity and $A_n$ denotes the normalization factor. For the modes we choose, $\boldsymbol{\mathcal{G}}_0(x) = A_0\cos(9\pi x/L)\boldsymbol{y}$ and $\boldsymbol{\mathcal{G}}_1(x) = A_1\cos(11\pi x/L)\boldsymbol{y}$, where $A_0=1.32\times10^{-5}$ [$V/\mu m$ ] and $A_1=1.46\times10^{-5}$  [$V/\mu m$ ]  are evaluated following the normalization procedure above. The variances can be evaluated using Eq.~\eqref{Eq:SIGMA_eval}
\begin{align}
    \sigma_0^2
    &=\frac{|\mu_0|^2A_0^2}{6}\left[1+\sinc(9r)\right] \\
    \sigma_1^2
    &=\frac{|\mu_0|^2A_1^2}{6}\left[1+\sinc(11r)\right]
\end{align}
where $\sinc(x)\equiv \sin(\pi x)/(\pi x)$ is the normalized sinc function.
The correlation can be evaluated using Eq.~\eqref{Eq:XI_eval}
\begin{equation}\label{Eq:XI_idealmirror}
    |\xi_{01}(r)| = \frac{| \sinc(10r) + \sinc(r)|}{\sqrt{ \left( 1+ \sinc(9r) \right) \left( 1+ \sinc(11r) \right) }}
\end{equation}
suggesting that the beating behavior of the two modes is the root cause of the oscillatory trend in the $\xi$--$r$ curve in Fig.~\ref{fig:xi_all} (a).

By plotting the numerator and the denominator of the equation above in Fig.~\ref{fig:xi_theory_kernel}, one can see that the numerator consists of a low-frequency term and a high-frequency term; the low-frequency term explains the increasing trend, while the high-frequency term explains the oscillatory behavior. The denominator is also oscillatory but the main effect is it regulates the curve so that it is bounded from 0 to 1. 

\begin{figure}[htbp]
    \centering
    \includegraphics[width=0.5\linewidth]{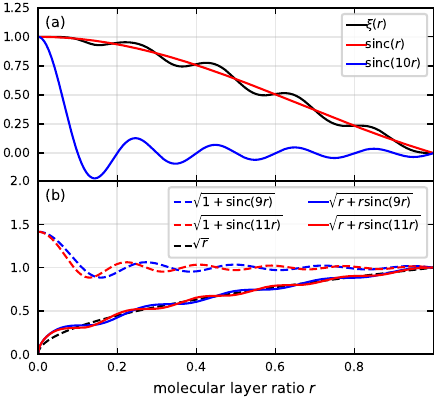}
    \caption{\label{fig:xi_theory_kernel}
    Analytical decomposition of the mode parameters for the ideal-mirror cavity configuration as a function of the molecular layer ratio $r$. 
    (a) Decomposition of the multi-mode correlation factor $\xi(r)$ for the ideal-mirror scenario. The underlying beating behavior is governed by the interplay between a low-frequency envelope component, $\text{sinc}(r)$ (red curve), and a high-frequency carrier component, $\text{sinc}(10r)$ (blue curve). The denominator acts as an oscillatory regulator, bounding the total correlation factor $\xi$ between 0 and 1.
    (b) Analytical profiles of the standard deviations $\sigma_0(r)$ and $\sigma_1(r)$ for the ideal-mirror case (Case M). The dashed curves represent the unscaled components $\sqrt{1 + \text{sinc}(9r)}$ (blue) and $\sqrt{1 + \text{sinc}(11r)}$ (red). When multiplied by the scaling factor $\sqrt{r}$ (black dashed curve), they yield the effective couplings $\sqrt{r}\sigma_k(r)$ (solid curves). The combination of the oscillatory $\sigma_1$ component and the monotonically increasing $\sqrt{r}$ factor explains the overall upward trajectory observed in Fig. 1(b).
    }
\end{figure}

\section{Collective mode expansion under frequency disorder}\label{appendix:binomial_freq_disorder}

\subsection{Perturbation treatment}\label{appendix:freq_disorder}
In the presence of frequency disorder in molecules, we present a perturbation analysis of the collective mode up to the second order. The perturbed total Hamiltonian takes the form $ \hat{H} = \hat{H} + \lambda\hat{V}'$ where the perturbation operator is $\hat{V}'= \sum_{j=1}^N \hbar\delta\omega_j \hat{c}_j^\dagger \hat{c}_j$. We assume the frequency disorder amplitude $|\delta\omega_j|$ to be small relative to $\Gamma_v/\hbar$ and $\omega_j$ is uncorrelated to the emitter positions or orientations.

One can understand the effect of frequency disorder $\hat{V}'$ by evaluating the Heisenberg equation of motion of the collective modes
\begin{equation}
    \hbar(\Omega_k - i\partial_t)\hat{a}_{k} + \sqrt{N}\bar{u}_k^*\hat{B}+\sqrt{N}\sigma_{k}\hat{D}_{k}=-g_{k}\mathcal{E}(t)
\end{equation}

\begin{equation}
    \hbar(\Omega_v-i\partial_t)\hat{B}+ \sum_{k'=1}^M\sqrt{N}\bar{u}_{k'}\hat{a}_{k'}+\frac{\lambda\hbar}{\sqrt{N}}\sum_{j=1}^N\delta\omega_j\hat{c}_j=0
\end{equation}

\begin{equation}
        \hbar(\Omega_v-i\partial_t)\hat{D}_{k} +\sum_{k'=1}^M\xi_{k,k'}\sqrt{N}\sigma_{k'}\hat{a}_{k'}
        + \frac{\lambda\hbar}{\sqrt{N}\sigma_k}\sum_{j=1}^N\delta\omega_j\delta u_{jk}^*\hat{c}_j=0
    \label{eq:dark_EOM_time_freqdis}
\end{equation}
Here we define the fluctuation mode $\hat{F}\equiv\frac{1}{\sqrt{N}\sigma_\omega}\sum_j\delta\omega_j\hat{c}_j$ and the auxiliary modes $\hat{G}_k\equiv\frac{1}{\sqrt{N}\sigma_k\sigma_\omega}\sum_j\delta\omega_j\delta u_{jk}^*\hat{c}_j$.
We then expand the operators $\hat{B}$, $\hat{D}_k$, $\hat{c}_j$, and $\hat{a}_k$ in power series of $\lambda$. 
\begin{align} 
\hat{c}_j &=\hat{c}_j^{(0)} +\lambda \hat{c}_j^{(1)}+\mathcal{O}(\lambda^2) \\
\hat{a}_k &= \hat{a}_k^{(0)} + \lambda\hat{a}_k^{(1)}+\lambda^2\hat{a}_k^{(2)}+\mathcal{O}(\lambda^3) \\
\hat{B} &= \hat{B}^{(0)} + \lambda\hat{B}^{(1)}+\lambda^2\hat{B}^{(2)}+\mathcal{O}(\lambda^3) \\
\hat{D}_k &= \hat{D}_k^{(0)} + \lambda\hat{D}_k^{(1)}+\lambda^2\hat{D}_k^{(2)}+\mathcal{O}(\lambda^3) \\
\hat{F} &= \hat{F}^{(0)} + \lambda\hat{F}^{(1)}+\mathcal{O}(\lambda^2) \\
\hat{G}_k &= \hat{G}_k^{(0)} + \lambda\hat{G}_k^{(1)}+\mathcal{O}(\lambda^2) 
\end{align}

The first-order perturbation of $\hat{c}_j$ satisfies $\hbar(\Omega_v-i\partial_t)\hat{c}_j^{(1)} + \hbar\delta\omega_j\hat{c}_j^{(0)}=0$. The first order terms of $\hat{a}_k^{(1)}$, $\hat{B}^{(1)}$, and $\hat{D}_k^{(1)}$ depend only on the zeroth order of $\hat{F}^{(0)}$ and $\hat{G}_k^{(0)}$.
\begin{equation}
    \hbar(\Omega_k - i\partial_t)\hat{a}_{k}^{(1)} + \sqrt{N}\bar{u}_k^*\hat{B}^{(1)}+\sqrt{N}\sigma_{k}\hat{D}_{k}^{(1)}=0
\end{equation}
\begin{equation}
    \hbar(\Omega_v-i\partial_t)\hat{B}^{(1)}+ \sum_{k'=1}^M\sqrt{N}\bar{u}_{k'}\hat{a}_{k'}^{(1)}+\hbar\sigma_\omega\hat{F}^{(0)}=0
\end{equation}
\begin{equation}
    \hbar(\Omega_v-i\partial_t)\hat{D}_{k}^{(1)} +\sum_{k'=1}^M\xi_{k,k'}\sqrt{N}\sigma_{k'}\hat{a}_{k'}^{(1)} 
        +\hbar\sigma_\omega\hat{G}_k^{(0)}=0
\end{equation}
Because $\delta\omega_j$ and $\hat{c}_j^{(0)}$ are uncorrelated, we have $\hat{F}^{(0)}=0$. Similarly, because $\delta\omega_j$ and $\delta u_{jk}^*\hat{c}_j^{(0)}$ are uncorrelated, we have $\hat{G}_k^{(0)}=0$. We find that the equations of motion for the first-order terms turn out to be identical with Eq.~(\ref{eq:photon_EOM_time}--\ref{eq:bright_EOM_time}) without the external drive, thus $\hat{a}_k^{(1)}$, $\hat{B}^{(1)}$, and $\hat{D}_k^{(1)}$ can be set to zero.

The second-order terms of the operators $\hat{a}_k^{(2)}$, $\hat{B}^{(2)}$, and $\hat{D}_k^{(2)}$ follow:
\begin{align}
    &\hbar(\Omega_k - i\partial_t)\hat{a}_{k}^{(2)} + \sqrt{N}\bar{u}_k^*\hat{B}^{(2)}+\sqrt{N}\sigma_{k}\hat{D}_{k}^{(2)}=0\\
    &\hbar(\Omega_v-i\partial_t)\hat{B}^{(2)}+ \sum_{k'=1}^M\sqrt{N}\bar{u}_{k'}\hat{a}_{k'}^{(2)}+ \hbar\sigma_\omega \hat{F}^{(1)}=0
    \\
    &\hbar(\Omega_v-i\partial_t)\hat{D}_{k}^{(2)} +\sum_{k'=1}^M\xi_{k,k'}\sqrt{N}\sigma_{k'}\hat{a}_{k'}^{(2)} 
    + \hbar \sigma_\omega\hat{G}_k^{(1)}=0 
\end{align}
in which the first-order terms of $\hat{F}^{(1)}$ and $\hat{G}_{k}^{(1)}$ are non-zero following
\begin{align}
    &\hbar(\Omega_v-i\partial_t)\hat{F}^{(1)} + \hbar\sigma_\omega\hat{B}^{(0)}=0
    \\
    &\hbar(\Omega_v-i\partial_t)\hat{G}_{k}^{(1)} + \hbar\sigma_\omega \hat{D}_k^{(0)}=0
\end{align}
Next, one can further simplify the equations of motion by letting $\bar{u}_k=0$ due to orientation disorder. The equations of motion for the second-order operators are 
\begin{align}
    &\hbar(\Omega_k - i\partial_t)\hat{a}_{k}^{(2)} + \sqrt{N}\sigma_{k}\hat{D}_{k}^{(2)}=0\\
    &\hbar(\Omega_v-i\partial_t)\hat{D}_{k}^{(2)} +\sum_{k'=1}^M\xi_{k,k'}\sqrt{N}\sigma_{k'}\hat{a}_{k'}^{(2)} 
    + \hbar \sigma_\omega\hat{G}_k^{(1)}=0 
    \\
    &\hbar(\Omega_v-i\partial_t)\hat{G}_{k}^{(1)} + \hbar\sigma_\omega \hat{D}_k^{(0)}=0
\end{align}
suggesting that the dark mode $\hat{D}_k$ exchanges energy with the auxiliary mode $\hat{G}_k$. 
These equations of motion can be solved in the frequency domain. Under a driving frequency $\omega$, the steady-state amplitudes satisfy the following equations:
\begin{align}
    &\hbar(\Omega_k - \omega)\ov{a}_{k}^{(2)} + \sqrt{N}\sigma_{k}\ov{D}_{k}^{(2)}=0\\
    &\hbar(\Omega_v-\omega)\ov{D}_{k}^{(2)} +\sum_{k'=1}^M\xi_{k,k'}\sqrt{N}\sigma_{k'}\ov{a}_{k'}^{(2)} 
    + \hbar \sigma_\omega\ov{G}_k^{(1)}=0 
    \\
    &\hbar(\Omega_v-\omega)\ov{G}_{k}^{(1)} + \hbar\sigma_\omega \ov{D}_k^{(0)}=0
\end{align}
We then conclude that the dark state population is contributed by both the dark modes and the auxiliary modes, and we can approximate the dark state population by $\tilde{P}_k\equiv |\ov{D}_k^{(0)} +\ov{D}_k^{(2)}|^2  + |\ov{G}_k^{(1)}|^2$.

In Fig.~\ref{fig:one_dimensional_all_focus}, we show that the emerging peak at $\omega_v$ is solely contributed by the auxiliary mode $\ov{G}_k^{(1)}$ with the rest of the peaks unchanged. This suggest that under small frequency disorder as a perturbation, DIMA is mostly unaffected, where the effect of frequency disorder manifests as a peak at $\omega_v$ in the dark state population.

\begin{figure}[htbp]
    \centering
    \includegraphics[width=0.5\linewidth]{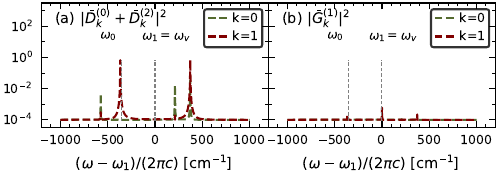}
    \caption{
    Steady-state amplitudes as a function of driving frequency $\omega$ for (a) the dark mode $|\ov{D}_k^{(0)} +\ov{D}_k^{(2)}|^2$ and (b) the auxiliary mode $|\ov{G}_k^{(1)}|^2$. The cavity setting is the same as Fig.~\ref{fig:one_dimensional_all}.  Note that the emergence of the dark state population at $\omega_v$ is solely contributed by the auxiliary mode $\bar{G}_k$ due to frequency disorder. }
    \label{fig:one_dimensional_all_focus}
\end{figure}

\subsection{Binary Frequency Disorder}
In order to model frequency disorder, here we consider a molecular ensemble in which the two-level systems have two possible frequencies, denoted by $\omega_{j} = \omega_v+\delta\omega_j$ where $\delta\omega_j=\pm\sigma_\omega$. We assume equal probability for $+$ and $-$, so that the average frequency remains $\omega_v$ and the standard deviation of the frequency distribution is $\sigma_\omega$. We can thus collect the molecule $j$ into two subsets $\nu=\{j|\delta\omega_j =\pm\sigma_\omega\}$ and define the corresponding symmetric modes $\hat{B}_\nu=\frac{1}{\sqrt{N/2}}\sum_{j\in\nu}\hat{c}_j$ and dark mode $\hat{D}_{k,\nu}=\frac{1}{\sigma_{k}\sqrt{N/2}}\sum_{j\in\nu}\delta{u}_{jk}^*\hat{c}_{j}$. One can follow the same procedure as in the main text and arrive at the steady-state equations
\begin{align}
    &\hbar(\Omega_k - \omega)\ov{a}_{k} + \sum_\nu\sqrt{\frac{N}{2}}\ov{u}_k^*\ov{B}_\nu+\sqrt{\frac{N}{2}}\sigma_{k}\ov{D}_{k,\nu}=-g_{k}\mathcal{E}_0 \label{eq:Ak_EOM_steady_binomial}\\
    &\hbar(\Omega_\nu-\omega)\ov{D}_{k,\nu}+\sum_{k'}\xi_{k,k'}\sqrt{\frac{N}{2}}\sigma_{k'}\ov{a}_{k'}=0 \label{eq:Dk_EOM_steady_binomial}\\
    &\hbar(\Omega_\nu-\omega)\ov{B}_\nu + \sum_{k'}\sqrt{\frac{N}{2}}\ov{u}_k\ov{a}_k=0\label{eq:bright_EOM_steady_binomial}
\end{align}
$\ov{B}_\nu$ can be omitted when  $\ov{u}_k=0$.
We then calculate the dark state population corresponding to mode $k$ as $P_k^{(\text{binary})} \equiv\sum_\nu|\ov{D}_{k,\nu}|^2$. As one can see $\sum_\nu|\ov{D}_{k,\nu}|^2\rightarrow|\ov{D}_{k}|^2$ as $\sigma_\omega\rightarrow0$ .

\subsubsection{Small Perturbation regime}
In terms of the auxiliary mode induced by frequency disorder, we find that the dark state population satisfies the following relationship under binary frequency disorder:
\begin{align}
P_k^{(\text{binary})} &= |\ov{D}_{k,+}|^2 + |\ov{D}_{k,-}|^2 \\
                                    &= |\frac{\ov{D}_k+\ov{G}_k}{\sqrt{2}}|^2 + |\frac{\ov{D}_k-\ov{G}_k}{\sqrt{2}}|^2\\
                                    &= |\ov{D}_k|^2 + |\ov{G}_k|^2
\end{align}
When evaluated within perturbation theory, we truncate the steady-state amplitudes of $\ov{D}_k$ to the second order, i.e. $\ov{D}_k \approx \ov{D}_k^{(0)} +\lambda^2\ov{D}_k^{(2)} + \mathcal{O}(\lambda^3)$ and keep $\ov{G}_k$ up to the first order term, i.e. $\ov{G}_k \approx \lambda\ov{G}_k^{(1)} + \mathcal{O}(\lambda^2)$. Substituting these truncated forms into the above relationship introduces a leading uncaptured cross-term error of order $\mathcal{O}(\lambda^3)$:
\begin{equation}
P_{k}^{(\text{binary})} = \tilde{P}_{k}^{(\text{binary})} + \mathcal{O}(\lambda^3)
\end{equation}
This shows that the perturbation theory accounts for the dark state population up to the second order $\mathcal{O}(\lambda^2)$, which explains the spectral agreement observed between Figs.~\ref{fig:one_dimensional_all_2}(b) and \ref{fig:one_dimensional_all_2}(d).

\subsubsection{Large Disorder Regime}
In order to understand how the perturbation theory breaks down for DIMA under large frequency disorder. Here we present the same spectra as Fig.~\ref{fig:one_dimensional_all_2} but with frequency disorder $\sigma_\omega= 50~[\text{cm}^{-1}]$,
\begin{figure}[htbp]
    \centering
    \includegraphics[width=0.5\linewidth]{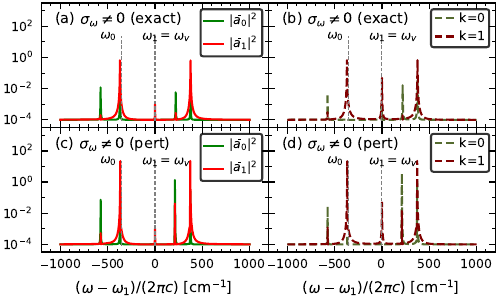}
    \caption{
    Steady-state spectra in the presence of large frequency disorder $\sigma_\omega=50$ [cm$^{-1}$]. Left column: The intensity of cavity photon modes ($|\bar{a}_0|^2$, $|\bar{a}_1|^2$) obtained by (a) exact calculation and (b) perturbation theory. Right column: The dark state population calculated by (b) $P_{k}^{(\text{binary})}=\sum_\nu|\bar{D}_{k,\nu}|^2$ and (d) perturbative solution  $\tilde{P}_{k}^{(\text{binary})}= |\bar{D}_k^{(0)} +\bar{D}_k^{(2)}|^2  + |\bar{G}_k^{(1)}|^2$.
    The cavity setting is the same as Fig.~\ref{fig:one_dimensional_all}. 
    The quantitative difference  between (a-b) and (c-d) shows the breakdown of perturbation theory at large $\sigma_\omega$.}
    \label{fig:one_dimensional_all_3}
\end{figure}

As one can see in Fig.~\ref{fig:one_dimensional_all_3}(b) the peak at $\omega_v$ is increased compared to the case in Fig.~\ref{fig:one_dimensional_all_2}(b). This is because the larger the frequency disorder term, the more mixing between the dark modes and auxiliary modes. Also, one can see that in Fig.~\ref{fig:one_dimensional_all_3}(a) there is a peak in $|\ov{a}_0|^2$  mode at $\omega_v$, which is not presented in the small $\sigma_\omega= 5~[\text{cm}^{-1}]$ case. While the qualitative features are similar, the peak amplitudes show quantitatively difference for the exact calautions (a-b) and the perturbative results (c-d). This suggests that higher-order perturbation is needed in the large $\sigma_\omega$ case.

\end{document}